\documentclass[11pt]{article}
\usepackage{graphicx}
\usepackage{epstopdf}
\usepackage{color}
\usepackage{xcolor}
\usepackage{epsfig,amssymb,amsmath}
  mm\evensidemargin=-5true mm \topmargin=-15true mm

\usepackage[
      colorlinks=true,
      linkcolor=blue,
      urlcolor=blue,
      filecolor=blue,
      citecolor=red,
      pdfstartview=FitV,
      pdftitle={},
      pdfauthor={},
      pdfsubject={},
      pdfkeywords={},
      pdfpagemode=None,
      bookmarksopen=true
]{hyperref}

\newtheorem{lemma}{Lemma}

\definecolor{FD}{rgb}{0.65,0.0,0}
\definecolor{geo}{rgb}{0,0.0,0.65}
\definecolor{GC}{rgb}{0,0.0,0.65}

  \newcommand{\rom}[1]{\mathrm{#1}}

  \newcommand{\be}{\begin{equation}}
  \newcommand{\ee}{\end{equation}}
  \newcommand{\bea}{\begin{eqnarray}}
  \newcommand{\eea}{\end{eqnarray}}
  
    \newcommand{\eps}{\epsilon}
  
  \newcommand{\p}{\partial}
  
  \newcommand{\s}{\sigma}
  
  \newcommand{\nn}{\nonumber}
\newcommand{\curl}{\mbox{curl}\,}
\newcommand{\sech}{\,\text{sech}\,}

  \newcommand{\n}{\nabla }

\newcommand{\beq}{\begin{equation}}
\newcommand{\eeq}{\end{equation}}
\newcommand{\beqs}{\begin{eqnarray}}
\newcommand{\eeqs}{\end{eqnarray}}

\newcommand{\Q}{{\cal Q}}
\newcommand{\DD}{{\mathcal{D}}}
\newcommand{\D}{{\nabla}}

\newcommand{\g}{\hspace{1pt}\mbox{}^3\hspace{-2pt}g}

\renewcommand{\th}{\theta}

\begin{document}

\begin{flushright}
\today
\end{flushright}

\begin{center}
{\Large \bf
Relaxing the Parity Conditions of Asymptotically Flat Gravity}
% No parity conditions for
%Asymptotically flat Einstein gravity without parity conditions:  \\ \mbox{} \vspace{-12pt}\\
%The Translation Anomaly}
 \vspace*{0.5cm}\\
 {Geoffrey Comp\`ere$^{\flat}$, Fran\c cois Dehouck$^{\natural}$
 }
\end{center}
 \vspace*{0.1cm}

 \begin{center}
 $^{\flat}${\it
 KdV Institute for Mathematics and Instituut voor Theoretische Fysica \\
Universiteit van Amsterdam, The Netherlands
 }
 \vspace*{0.5cm}\\
 $^{\natural}$$^{\flat}${\it
 Physique Th\'eorique et Math\'ematique,  \\ Universit\'e Libre de
     Bruxelles and International Solvay Institutes, Bruxelles,  Belgium
 }
 \vspace*{0.5cm}\\
{\tt gcompere@ulb.ac.be} $\quad$  {\tt fdehouck@ulb.ac.be}

 \end{center}

\bigskip

\begin{abstract}
{\normalsize
Four-dimensional asymptotically flat spacetimes at spatial infinity are defined from first principles without imposing parity conditions or restrictions on the Weyl tensor. The Einstein-Hilbert action is shown to be a correct variational principle when it is supplemented by an anomalous counter-term which breaks asymptotic translation, supertranslation and logarithmic translation invariance. Poincar\'e transformations as well as supertranslations and logarithmic translations are associated with finite and conserved charges which represent the asymptotic symmetry group. Lorentz charges as well as logarithmic translations transform anomalously under a change of regulator. Lorentz charges are generally non-linear functionals of the asymptotic fields but reduce to well-known linear expressions when parity conditions hold. We also define a covariant phase space of asymptotically flat spacetimes with parity conditions but without restrictions on the Weyl tensor. In this phase space, the anomaly plays classically no dynamical role. Supertranslations are pure gauge and the asymptotic symmetry group is the expected Poincar\'e group.
}
\vspace{12pt}

\vspace{12pt}
 \end{abstract}

\newpage

\tableofcontents

\newpage

\section{Two puzzles in asymptotically flat spacetimes}

 Both in Hamiltonian framework or in covariant phase space methods, asymptotically flat spacetimes at spatial infinity have only been defined when parity conditions on the first order part of the boundary fields are imposed. These conditions have been introduced in \cite{Regge:1974zd} in the Hamiltonian formalism and later in the covariant phase space, see \cite{ABR}, so that Lorentz charges are finite and so that the canonical structure, or in Lagrangian formalism the covariant symplectic structure, is also finite. Now, it was noticed in \cite{Mann:2005yr} (see also earlier work \cite{GH,Ashtekar:1978zz,BY,HH,Kraus:1999di, Mann:1999pc, deHaro:2000wj,Astefanesei}) that asymptotically flat spacetimes at spatial infinity admit a variational principle whether or not parity conditions on the first order part of the boundary fields in the asymptotic cylindrical or hyperbolic radial expansion hold, at least when one neglects boundary terms at the past and future boundaries.

It is then intriguing that even though the action is finite, the symplectic structure and conserved charges are infinite when parity conditions do not hold. One might think that the action determines the entire dynamics and therefore in particular the symplectic structure and conserved charges but this expectation does not seem to be realized. This constitutes the first puzzle that we will resolve in this work\footnote{ We thank D. Marolf and A. Virmani for drawing our attention to this issue and A. Ashtekar for emphasizing the role of past and future boundary terms in the variational principle. A hint that parity conditions might not be required in a phase space including counter-terms was provided in  \cite{Compere:2011db} where it was shown that counter-term charges \cite{Mann:2005yr,Mann:2006bd,Mann:2008ay} and Ashtekar-Hansen charges \cite{Ashtekar:1978zz} are equivalent without using parity conditions, apart for the four lowest harmonics of the first order field to fix logarithmic translations,  in a restricted phase space where supertranslations are also completely fixed.}.

Asymptotically flat spacetimes are defined by boundary conditions at infinity. The class of diffeomorphisms which preserve the boundary conditions are the allowed diffeomorphisms. Allowed diffeomorphisms associated with non-trivial conserved charges - the large diffeomorphisms -  modulo diffeomorphisms associated with zero charges - the pure gauge transformations -  define the asymptotic symmetry group. For asymptotically flat spacetimes at spatial infinity when parity conditions are imposed, the asymptotic symmetry group has been worked out both with Hamiltonian and covariant phase space methods. In Hamiltonian framework, it is just  the Poincar\'e group. In that framework, parity odd supertranslations also preserve the boundary conditions but they are associated with vanishing Hamiltonian generators \cite{Regge:1974zd}. In covariant phase space, the asymptotic symmetry group has been shown to be the Poincar\'e group only in the truncated phase space where part of the first order fields have been set to zero, by setting the first order magnetic part of the Weyl tensor to zero. This last condition also fixes all supertranslations \cite{Ashtekar:1978zz,Ashtekar:1984aa}. The equivalence of Hamiltonian and Lagrangian formalisms tells us that there should exist a covariant phase space which allows for more generic first order fields (see also comments in \cite{Beig:1987aa}) and such that the covariant equivalent of the Hamiltonian odd-supertranslations act as pure gauge transformations. We will show that indeed such a generalization of the covariant phase space with parity conditions exists.

The existence of such a consistent covariant phase space or canonical phase space does not shed a complete light on the choice of boundary conditions. In particular, it is not clear whether or not covariant phase spaces related to each other by unallowed diffeomorphisms are inequivalent. This constitutes our second puzzle. Both in the usual covariant phase space and in Hamiltonian framework, different logarithmic translations or generic parity supertranslations are not transformations associated with finite Hamiltonian or Lagrangian generators when one uses the standard canonical bracket or symplectic structure. There is therefore no covariant phase space or canonical space that encompass spacetimes or initial data surfaces with such different asymptotic transformations. Now, either these transformations are unphysical or they are physical. On the one hand, if they are unphysical, one would expect that they are pure gauge (see \cite{AshLog} for arguments that logarithmic translations are unphysical). The fact that logarithmic translations and generic parity supertranslations are not degenerate directions of the symplectic structure - they are not allowed directions of the symplectic structure in the first place - is however in tension with the intuition that pure gauge transformations should be degenerate directions of the symplectic structure. On the other hand, if these transformations have a physical content, there should be a way to regularize the infinities. If an enlarged phase space exists where both logarithmic and generic supertranslations are allowed in the first place, it would allow to settle these questions. However, there does not exist in the literature a construction of a consistent phase space where parity conditions have not been imposed.

These two puzzles have in common that they rely on the standard covariant phase space symplectic structure or in Hamiltonian formalism on the canonical bracket defined from the canonical fields $\g_{ij}$ and $\pi^{ij}$. Now, it is important to remember that  the algebraic derivation of the covariant phase space symplectic structure from the bulk Lagrangian suffers from ambiguities \cite{Wald:1993nt,Iyer:1994ys}. Also, the definition of the canonical structure depends on what fields are considered to be canonical. For example, in \cite{Regge:1974zd}, it was pointed out that the asymptotic values for the shift and lapse functions should be considered as additional canonical variables in addition to $\g_{ij}$ and $\pi^{ij}$. The consideration of additional canonical variables might then lead to a modification of the canonical structure. It is also important to observe that in the arguments of \cite{Mann:2005yr} the variational principle is defined only under the assumption that future and past boundary terms can be dealt with without affecting the analysis at spatial infinity.

\section{Main results and outline}

The key  technical result in this paper is the fixation of ambiguities in the symplectic structure and the canonical bracket. We show that the ambiguities can be fixed from first principles by requiring the existence of a variational principle taking into account boundary terms at past and future infinity in addition to spatial infinity. Our definition of symplectic structure then amounts to the prescription given in \cite{CompereMarolf} which was developed following the counter-term methods \cite{Henningson:1998gx,Henningson:1998ey,Balasubramanian:1999re,de Haro:2000xn,Skenderis:2000in,deHaro:2000wj} in the framework of the AdS-CFT correspondence \cite{Maldacena:1997re,Witten:1998qj}. An important difference with earlier treatments of asymptotically flat spacetimes is the inclusion of a translation breaking counter-term, a translation anomaly, in the action. As a consequence of the addition of boundary counter-terms to the canonical structure and symplectic structure, no parity conditions are  required at any step in the construction of the phase space and finite conserved charges can be defined from first principles. As a result, supertranslations and logarithmic translations are associated with non-trivial conserved charges. These conserved charges are trivial when parity conditions hold.  Our Poincar\'e conserved charges reduce to the standard ADM \cite{Arnowitt:1962aa,Regge:1974zd} or Lagrangian charges \cite{Geroch:1977aa,Ashtekar:1978zz,AshRev,Abbott:1981ff,Ashtekar:1991vb, Wald:1993nt,Iyer:1994ys,Barnich:2001jy} when parity conditions hold. 
 We also define a covariant phase space of asymptotically flat spacetimes with parity conditions that generalizes previous constructions \cite{ABR,Ashtekar:2008jw}. In this phase space, the anomaly does not contribute to the symplectic structure so most of the conventional properties of asymptotically flat spacetimes are kept, e.g. the Poincar\'e group is the asymptotic symmetry group. These and other consequences of our construction will be addressed in the following sections.

We start in Section \ref{sec:BC} by providing the detailed boundary conditions we will use throughout the main text. We make connection with previous choices of boundary conditions and indicate how they are generalized. In Section \ref{sec:L}, we provide a first-principle derivation of the asymptotic dynamics of Einstein gravity, with our flat boundary conditions, starting from the action principle and deriving the asymptotic equations of motion, the symplectic structure, the covariant phase space charges and the asymptotic symmetry group. In Section \ref{sec:H} we present the main milestones on how our construction fits in the Hamiltonian framework. We end with a summary and detailed discussion in Section \ref{sec:D}. Some additional details on the boundary conditions and the equations of motion are relegated to appendices. We provide in Appendix \ref{app:BC} a comparison between covariant and 3+1 boundary conditions. In Appendix \ref{sec:SD} we obtain a classification of specific tensor fields useful to characterize the algebra of tensors at second order in the asymptotic expansion when all first order fields are considered. In Appendix \ref{appEOM} we derive the asymptotic equations of motion in the hyperbolic representation of spatial infinity up to second order in the radial expansion.

We refer the reader to the companion paper \cite{Compere:2011db} for a review of the hyperbolic representation of spatial infinity, our notations, and the presentation and proofs of some lemmas and properties of tensor fields on the hyperboloid. Although this paper deals only with smooth fields, a supplementary lemma on singular tensor fields on the hyperboloid relevant to describe NUT charges is stated in Appendix \ref{app:Nut} for completeness.

\section{Specification of boundary conditions}
\label{sec:BC}

The boundary conditions are part of the specification of the theory.  If the boundary conditions are too strong, there is no interesting physics in the restricted phase space. If they are too loose, no phase space can be defined at all because physical quantities like the total energy or the angular momenta diverge. Starting from a set of boundary conditions, one can try to enlarge them such that all physically reasonable solutions of the equations of motion lie inside the phase space and such that all physically relevant quantities like action, symplectic form and conserved charges are well-defined. Note that there is no theorem guaranteeing that this procedure leads to a unique ``largest'' phase space. In the sequel, we present  boundary conditions in covariant phase space and  3+1 formalism which are more general than the ones discussed for example in  \cite{ Regge:1974zd} and  \cite{ABR}. The consistency of these conditions will be discussed in the later sections. The relationship between boundary conditions in the canonical and covariant formalisms is further discussed in Appendix \ref{app:BC} following earlier comparisons \cite{Ashtekar:1984aa,Beig:1987aa}.

\subsection{Boundary conditions in the hyperbolic representation of spatial infinity}

In the hyperbolic representation of spatial infinity proposed in \cite{Ashtekar:1978zz} and developed in \cite{BS,B,Ashtekar:1991vb}, one demands that there should exist a coordinate system $(\rho,\tau,\theta,\phi)$ such that the four-metric takes the form
\bea
ds^2 = \Big(1+ \frac{2\s}{\rho} + o(\rho^{-1})\Big) d\rho^2 + O(\rho^0) \: d\rho \: dx^a + \Big( \rho^2 h_{ab}^{(0)} + \rho h_{ab}^{(1)}+ o(\rho^1) \Big) dx^a  dx^b \label{BC2} \: ,
\eea
where $h_{ab}^{(0)}$ is the unit hyperboloid
\bea
ds^2_{\mathcal{H}} &=& h_{ab}^{(0)}dx^a dx^b = -d\tau^2 + \cosh^2\tau (d\theta^2 + \sin^2\theta d\phi^2) \: ,
\eea
and $\s$ and $h^{(1)}_{ab}$ can be considered as a scalar and a tensor field defined on the hyperboloid. Minkowski spacetime takes this form when one transforms the standard spherical coordinates into $\rho^2 = \eta_{\mu\nu}x^\mu x^\nu = r^2 - t^2$ and $\tanh \tau = t/r$. We take as our covariant phase space all metrics that can be put into the form \eqref{BC2} and where $k_{ab}$, defined as
\bea
k_{ab} \equiv h_{ab}^{(1)} + 2 \s h_{ab}^{(0)},
\eea
obeys the constraints given below in \eqref{condk}. In the following, $\DD_a$ denotes the covariant derivative associated with the metric on the hyperboloid $h^{(0)}_{ab}$ and $\Box\equiv\DD^a \DD_a$. Indices are raised with $h^{(0)\: ab}$. 

As pointed out in \cite{BS,B}, the phase space is invariant under Poincar\'e transformations, supertranslations and logarithmic translations.
Logarithmic translations are defined as $\rho \rightarrow \rho + \log \rho \: H(x^a)  + o(\rho^0)$, $x^a \rightarrow x^a + \rho^{-1}(\log \rho +1)\:  \DD^a H(x^b) + o(\rho^{-1})$ where $H_{ab} + h_{ab}^{(0)} H = 0$\footnote{Our definition is equivalent to the linear combination of a logarithmic translation as defined in \cite{BS,B} with a translation.}. They act on the first order fields $\s$ and $k_{ab}$ as
\bea
\delta_H \s = H,\qquad  \delta_H k_{ab} = 0. \label{actlog}
\eea
The transformations with $H$ arbitrary are the so-called logarithmic supertranslations that could be used to set $\sigma = 0$ at the expense of introducing a logarithmic term of the form $\rho \log\rho (H_{ab} + h_{ab}^{(0)} H) dx^a dx^b$ in the metric. We choose to set this logarithmic term to zero in \eqref{BC2}  in order to follow the notation of previous literature. Logarithmic supertranslations are therefore not allowed diffeomorphisms with our choice of boundary conditions\footnote{Note that our second puzzle also applies to logarithmic supertranslations but for simplicity we will not consider these transformations in this work.}. 

Supertranslations are defined as the transformations $\rho \rightarrow \rho + \omega(x^a) + o(\rho^0)$, $x^a \rightarrow x^a + \rho^{-1} \DD^a \omega (x^b) + o(\rho^{-1})$. When acting on the leading order fields $\s$ and $k_{ab}$, we find
\bea
\delta_\omega \s = 0, \qquad \delta_\omega k_{ab} = 2 (\DD_a \DD_b \omega + \omega h_{ab}^{(0)}).\label{actsuper}
\eea
The trace of $k_{ab}$, $k \equiv  h^{(0)\: ab} \: k_{ab}$ is unconstrained by the equations of motion derived from Einstein's equations while the divergence of $k_{ab}$ has to obey $\DD^b k_{ab} = \DD_a k$. Supertranslations that change the trace $k$ are given by transformations \eqref{actsuper} that obey $(\square+3)\omega \neq 0$. As we will show in Section \ref{LagCharges}, one can associate charges to such transformations  that are not conserved at spatial infinity. Since we aim at defining a phase space with only conserved charges at spatial infinity, we will discard such transformations by fixing the trace of $k$ to an arbitrary scalar $\bar k$. Consistently with the equations of motion, we impose the boundary conditions that $k_{ab}$ has fixed trace and divergence 
\bea
k \equiv  h^{(0)\: ab} \: k_{ab} = \bar k,\qquad  \DD^b k_{ab} = \DD_a \bar k. \label{condk}
\eea
For simplicity, we will fix $\bar k =0$. As we will see in Section~\ref{sec:action}, the conditions \eqref{condk} turn out to be sufficient in order that the Mann-Marolf action \cite{Mann:2005yr} be a variational principle on the phase space. Accordingly to the restriction \eqref{condk}, we restrict the function $\omega(x^a)$ on the hyperboloid to obey
\bea
(\square +3 )\omega = 0 \, ,\label{super}
\eea
which is a severe restriction with respect to the definition of supertranslations used e.g. in \cite{ABR}. Translations are singled-out as the four transformations obeying
\bea
\omega_{ab} + h_{ab}^{(0)}\omega = 0, \qquad \omega_{ab} \equiv \DD_a \DD_b \omega\, .\label{eq:tra}
\eea
By convention, we will refer to supertranslations as the transformations generated by $\omega$ which obey \eqref{super} but do not obey \eqref{eq:tra}. 

In order to make connection with previous treatments, let us review what the boundary conditions of \cite{Ashtekar:1978zz} impose in addition to \eqref{BC2}. As we review in Appendix \ref{appEOM}, once fixing $k=0$, Einstein's equations imply that
\bea
\Big(\curl (\curl k) \Big) _{ab} =(\Box-3)k_{ab}= 0 ,\qquad (\curl T)_{ab} \equiv \eps_{a}^{\;\,  cd} \DD_c T_{db}\, .
\eea
It has been shown in \cite{Ashtekar:1978zz} (referred to as Lemma 1 in \cite{Compere:2011db}) that these equations imply that $k_{ab}$ is determined in terms of a scalar $\beta$ as
\bea
\curl k_{ab} = \DD_a \DD_b \beta + h_{ab}^{(0)} \beta \: ,
\eea
with $(\square +3)\beta = 0$. In \cite{Ashtekar:1978zz}, it is imposed that
\bea
\beta = 0.\label{BC2s}
\eea
Once the curl of $k_{ab}$ is zero, one can use again the Lemma in \cite{Ashtekar:1978zz} to express $k_{ab} = \DD_a \DD_b I + h_{ab}^{(0)}  I$ with $(\square+3)I=0$. One can then fix completely the supertranslations by imposing
\bea
k_{ab} = 0.\label{kab}
\eea

In addition to this, it can be imposed (see e.g. \cite{ABR} even thought this condition was not imposed in the original reference  \cite{Ashtekar:1978zz}) that $\sigma$ obeys the even parity-time reversal condition
\bea
\sigma(\tau , \theta, \phi ) = \sigma (-\tau , \pi - \theta ,\phi + \pi)\label{Par3} \:,
\eea
so that the standard covariant phase space symplectic structure is finite.  Remark that this condition also completely removes the freedom of performing logarithmic translations since $H$ are (parity-time) odd functions on the hyperboloid. In this work we will relax the conditions \eqref{BC2s}, \eqref{kab} and \eqref{Par3} but impose that $k_{ab}$ is traceless and divergence-free.

To derive the conserved charges associated with rotations and boosts, the specification of the fall-off conditions \eqref{BC2} are not sufficient. In this work, we will restrict ourselves to the phase space where the metric can be put in the following Beig-Schmidt form up to second order
\bea
\label{metric1}
ds^2 &=& \left( 1 + \frac{2\s}{\rho}+ \frac{\s^2}{\rho^2}  + o(\rho^{-2})\right) d \rho^2 +  o(\rho^{-1}) d\rho dx^a \nn\\
&& + \rho^2\left( h^{(0)}_{ab} + \frac{h^{(1)}_{ab}}{\rho} + \ln\rho \frac{i_{ab}}{\rho^2} + \frac{h^{(2)}_{ab}}{\rho^2} + o(\rho^{-2}) \right) dx^a dx^b .
\eea
The logarithmic term $i_{ab}$ is necessary and sufficient in order that the equations of motion have a solution in general, up to second order in the expansion, when $\s$ and $k_{ab}$ do not obey parity conditions (see Section \ref{sec:LS}). The set of metrics \eqref{metric1} is a consistent truncation of the phase space \eqref{BC2}, which is preserved by Poincar\'e transformations, logarithmic  translations and supertranslations. Note that subleading terms in those transformations are constrained such that the form \eqref{metric1} is preserved.

\subsection{Boundary conditions in Hamiltonian formalism}

It is generally accepted that asymptotically flat spacetimes in 3+1 formalism should be defined using the following fall-off conditions on the three-dimensional metric $\g_{ij}$ and its conjugate momentum $\pi^{ij}$
\bea
\g_{ij} &= & \delta_{ij}+O(r^{-1}), \qquad \qquad \;\;\qquad \partial_k \g_{ij} = O(r^{-2}), \\
\pi^{ij} &=& O(r^{-2}) \, ,\nn
\eea
on the initial data surface $\Sigma$ defined as $t=0$. Here Latin indices denote spatial components: $x^i = x^1,x^2,x^3$ in Cartesian coordinates and $r^2 = x_i \, x^i$. We will use capital Latin indices $A,B, \dots$ to denote both normal ($\perp$) and spatial components and greek letters $\iota, \zeta,\xi$ as indices on the two-sphere in spherical coordinates.

Following \cite{Regge:1974zd}, ten asymptotic values for the lapse and shift functions $N^A$, $M^{AB} = M^{[AB]}$, are introduced as additional canonical variables (together with their conjugate momenta) on the same footing as $\g_{ij}$, $\pi^{ij}$. They specify the asymptotic location of the initial time Cauchy surface $\Sigma$ at spatial infinity through the asymptotic values of the lapse $N$ and shift functions $N^i$
\bea
N &=& M^{\;\, \perp}_{i} x^i + N_\infty + O(r^{-1}), \qquad \partial_ i N = M^{\;\, \perp}_{i}  + O(r^{-2}) , \nn\\
N^i &=& M^{\; \,i}_j x^j + N^i_\infty +O(r^{-1}) ,\qquad \partial_j N^i = M^{\; \,i}_j + O(r^{-2}),\label{BC1bis}
\eea
where
\bea
M_{ij} =   M^{\; \, k}_i \delta_{kj}= - M_{ji},\qquad  M^{\;\, \perp}_i = -M_{i\perp} = M_{\perp i}\, .
\eea

In the original Hamiltonian treatment  \cite{Regge:1974zd}, the leading asymptotic parts of the canonical fields
\bea
\g_{ij} &= & \delta_{ij} + \frac{1}{r} \g_{ij}^{(1)} + o(r^{-1}) ,\label{geng1} \\
\pi^{ij} & = & \frac{1}{r^2} \pi^{(2)\: ij} + o(r^{-2}) , \label{geng2}
\eea
are restricted to obey the parity conditions
\bea
\g_{ij}^{(1)}(- \mathbf n) =  \g_{ij}^{(1)}( \mathbf n),\qquad \pi^{(2) \: ij} (-\mathbf n) = - \pi^{(2)\: ij}(\mathbf n) , \label{Par1}
\eea
where $\mathbf n = r^{-1}(x^1,x^2,x^3)$. These conditions have been shown to be necessary in order that the Hamiltonian defined from the canonical variables $\g_{ij}$, $\pi^{ij}$, $M^{AB}$, $N^A$ is finite when asymptotic rotations and boosts are considered. The full class of transformations preserving the fall-off and parity conditions is larger than the Poincar\'e group. Indeed, angle-dependent shifts of the lapse and shift functions
\bea
N^\perp  &=& S^\perp (\mathbf n ) + o(r^0),\qquad N^i  = S^i (\mathbf n ) + o(r^0),  \label{Hamsuper}
\eea
are allowed transformations once they are restricted to be parity odd
\bea
S^\perp (- \mathbf n) = -  S^\perp (\mathbf n),\qquad S^i (- \mathbf n) = -  S^i (\mathbf n) . \label{Par2}
\eea
We will refer to $S^\perp(\mathbf n)$ as temporal supertranslations, $S^r(\mathbf n)$ as radial supertranslations and supertranslations tangent to $r$ constant surfaces $S^\zeta (\mathbf n)$ as angular supertranslations.
%However, it was shown in \cite{Regge:1974zd} that these shifts do not affect the Poincar\'e charges.

In this work, we relax the parity conditions \eqref{Par1}. Once parity conditions are relaxed, the fall-off conditions are preserved by both parity even and parity odd supertranslations \eqref{Hamsuper}, i.e. conditions \eqref{Par2} are no longer required. In addition, the fall-off conditions on the asymptotic fields $\g_{ij}$ and $\pi^{ij}$ are preserved under the so-called logarithmic translations which are associated to the lapse and shift functions
\bea
N^\perp &=& \log r \, K^\perp + o(r^0) ,\\
N^i &=& \log r \, K^i + o(r^0) ,
\eea
where $K^\perp$ and $K^i$ are constants.

In order to derive the explicit formulae for the conserved Hamiltonian charges associated with rotations and boosts, it is required to specify the subleading components of the fields in \eqref{geng1}-\eqref{geng2}.  For this purpose, we restrict ourselves to the truncation of the canonical phase space where the fields obey
\bea
\g_{ij} &= & \delta_{ij} + \frac{1}{r} \g_{ij}^{(1)}+ \frac{\log r}{r^2} \g_{ij}^{(ln,2)} + \frac{1}{r^2} \g_{ij}^{(2)} + o(r^{-2}) ,\label{geng1sub} \\
\pi^{ij} & = & \frac{1}{r^2} \pi^{(2)\: ij}+ \frac{\log r}{r^3} \pi^{(ln,3)\: ij}+ \frac{1}{r^3} \pi^{(3)\: ij}+ o(r^{-3})\, .\label{geng2sub}
\eea
The logarithmic branch is necessary in order that the Hamiltonian and momentum constraints admit solutions in general when parity conditions \eqref{Par1} do not hold, as already noticed e.g. in Appendix B of \cite{Beig:1987aa}.

A comparison of boundary conditions between 3+1 cylindrical and four-dimensional hyperbolic representations can be made following \cite{Ashtekar:1984aa,Beig:1987aa}. We detail this comparison in Appendix \ref{app:BC}. Note that Poincar\'e and logarithmic translations can be mapped between 3+1 and covariant formalisms. There is also a one-to-one mapping between canonical temporal and radial supertranslations and covariant supertranslations verifying \eqref{super}. Covariant angular supertranslations generate mixed components $g_{\rho a}$ in hyperbolic asymptotic coordinates and are usually fixed in the Beig-Schmidt expansion \cite{BS}.
As our approach is based on the existence of a variational principle, we only consider covariant boundary conditions transposed to the Hamiltonian formalism. We will thus relax conditions \eqref{Par1} and \eqref{Par2} but impose additional boundary conditions on the fields such that the angular supertranslations are fixed. These last conditions impose that the a priori generic form of the fields present in \eqref{geng1sub}-\eqref{geng2sub} is actually fixed in terms of their covariant counterparts as detailed in Appendix \ref{app:BC}. It would be interesting, however out of the scope of this paper, to include mixed terms  $g_{\rho a}$ in the Beig-Schmidt expansion and allow for angular supertranslations along the lines presented in the next section\footnote{ In the presence of mixed terms $g_{\rho a}$, there might be a distinction between the bulk covariant phase space symplectic structure defined from the action and the one defined from the equations of motion, see definitions of $\omega$ and $W$ in Section \ref{sec:symplectic}. One would then need to prescribe which one is the bulk symplectic structure, see \cite{Azeyanagi:2009wf} for an example where such a prescription plays an important role.}.

\section{Lagrangian dynamics in the hyperbolic representation}
\label{sec:L}

\subsection{Action principle}
\label{sec:action}

In this section, we define the action principle for asymptotically flat spacetimes using hyperbolic temporal and spatial cutoffs. We introduce a finite hyperbolic cut-off $\rho = \Lambda$ in order to regulate spatial infinity. We denote as $\mathcal H$ the unit hyperboloid defined at the cut-off $\rho = \Lambda$. We also limit the domain where the variational principle is defined between an initial and a final hyperbolic spacelike hypersurface that we denote as $\Sigma_\pm$. Such temporal slices are relatively boosted with respect to each other close to spatial infinity. The spheres lying at the intersection of the boundary hyperboloid $\mathcal H$ with the hypersurfaces $\Sigma_\pm$ are denoted as $S_\pm$ and are defined at hyperboloid times $\tau = \tau_\pm$, see Figure~1.

\begin{figure}[!hbt]
  \centering \label{fig1}
 % \resizebox{0.30\textwidth}{!}{
\includegraphics[width=0.3\textwidth]{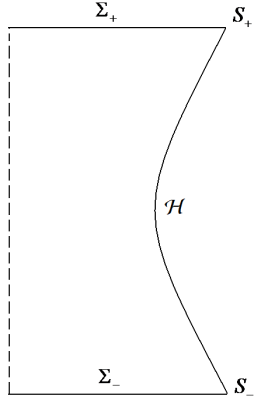}
%}
  \caption{The variational principle is defined in the spacetime delimited by initial and final hyperbolic temporal slices $\Sigma_\pm$ and the hyperbolic radial cut-off $\mathcal H$.}
\end{figure}

The notation we will use through Section \ref{sec:L} is as follows. We will denote the volume form as
\bea
(d^{n-p} x)_{\mu_1 \dots \mu_p}\equiv \frac{1}{p! (n-p)!} \epsilon_{\mu_1 \dots \mu_n} dx^{\mu_{p+1}} \wedge \dots \wedge dx^{\mu_n},
\eea
and we also have $\delta g^{\mu\nu}\equiv g^{\mu \alpha} g^{\nu \beta} \delta g_{\alpha \beta}$. The covariant derivative associated to $g_{\mu\nu}$ is $\nabla_{\mu}$. Two-forms are defined as $\mathbf{k}=\frac{1}{2} k_{\mu\nu} dx^{\mu} \wedge dx^{\nu}=\tilde{k}^{\mu\nu} (d^2 x)_{\mu\nu}$ implying that $d\mathbf{k}=\frac{1}{2} \n_\s k_{\mu\nu} dx^\s \wedge dx^{\mu} \wedge dx^{\nu}=\n_{\nu} \tilde{k}^{\mu\nu} (d^3 x)_{\mu}$. For Stokes' theorem to hold, we have set conventions that imply $\epsilon_{\tau \rho \theta \phi}=\epsilon_{\rho \theta \phi}=-\epsilon_{\tau \theta \phi}=\epsilon_{\theta \phi}$. This also tells us that $(d^3x)_{\tau}= -d\rho (d^2 x)_{\tau}$, $(d^3 x)_{\rho}=-d\tau (d^2 x)_{\rho}$  and $(d^2x)_{\tau}=-(d^2x)_{\rho}=2(d^2x)_{\rho\tau}=-d^2S$ where $d^2S=\frac{1}{2} \epsilon_{\zeta \iota} dx^{\zeta} \wedge dx^{\iota}$.

Using a regulation with hyperbolic cutoffs, the action for asymptotically flat spacetimes has the form 
\bea
S = \frac{1}{16\pi G} \int_{\cal M} d^4 x \sqrt{-g}\,R + \mathcal S_{\Sigma_\pm} + \mathcal S_{\mathcal H}\, , \label{covaction}
\eea
where $\mathcal S_{\Sigma_\pm}$ are boundary terms at the future and past boundaries and $\mathcal S_{\mathcal H} $ are boundary terms at the radial boundary. It is interesting to first concentrate our attention on the past and future boundaries that are usually not treated in the formulation of a variational principle with zero cosmological constant. Upon varying the action, the boundary terms on $\Sigma_\pm$ can be written as
\bea
\delta S |_{\Sigma_\pm} = \pm \frac{1}{16 \pi G}\int_{\Sigma_\pm} (d^3 x)_\mu \Theta^\mu [\delta g] + \delta \mathcal S_{\Sigma_\pm} + \mathcal B_{S_\pm} \, ,\label{vara1}
\eea
where
\bea
\Theta^\mu[\delta g] (d^3 x)_\mu = \sqrt{-g}\left( g^{\alpha\beta}\delta \Gamma_{\alpha \beta}^\mu - g^{\mu\alpha}\delta \Gamma_{\alpha\beta}^\beta \right)(d^3 x)_\mu
\eea
is the presymplectic form obtained from $\delta (\sqrt{-g}R) = - \sqrt{-g} G^{\mu\nu} \delta g_{\mu\nu} +\p_\mu \Theta^\mu$. The terms $\mathcal B_{S_\pm}$ defined on the boundary spheres $S_\pm$ are obtained from collecting the boundary terms in the variation of the action  $\mathcal S_{\mathcal H} $ with respect to boundary fields. Using the asymptotic expansion of the fields and $\int_{\Sigma} (d^3 x)_a \omega^a = - \int^{\rho = \Lambda} d\rho \int_S (d^2 x)_a \omega^a$, we obtain that  the first term in equation \eqref{vara1} contains a linearly divergent term, a logarithmic divergent term and a finite term $ \mathcal F_{\Sigma_\pm}$ that can be written as
\bea
\delta S |_{\Sigma_\pm} &=& \pm \frac{\log \Lambda}{16 \pi G} \int_{S_\pm} d^2 S \sqrt{-h_{(0)}}n_a \left( 4 \delta \s \DD^a \s + \frac{1}{2}\delta k^{bc} \DD^a  k_{bc}-\delta k^{bc}\DD_c k^a_{\; b} \right) \nn\\
&&+ \delta \left( \mathcal S_{\Sigma_\pm} +\log\Lambda \hat{\mathcal R}_{S_\pm}+ \Lambda \mathcal R_{S_\pm} \right) + \mathcal F_{\Sigma_\pm} + \mathcal B_{S_\pm} \, .  \label{varS2}
\eea
The finite term $\mathcal F_{\Sigma_\pm}$, the linearly divergent term  $\Lambda \delta \mathcal R_{S_\pm}$ as well as the logarithmically divergent term $\log\Lambda \hat{\mathcal R}_{S_\pm}$ are expected to be exactly canceled by an appropriate choice of counterterm $\mathcal S_{\Sigma_\pm}$ after an appropriate choice of boundary conditions at future and past times has been made. The derivation of the boundary conditions and boundary terms at $\Sigma_\pm$ would require a careful analysis that we will not perform here. Now, the logarithmic divergent term at $\Sigma_\pm$ cannot be canceled by a local algebraic expression of the boundary fields alone. Indeed, there is no logarithmic branch in the asymptotic expansion of the metric \eqref{BC2} at leading enough order in $\rho$. We will discuss two mechanisms to set that divergence to zero. 

One mechanism to cancel the logarithmic divergence is to impose the following parity conditions
\bea
\sigma(\tau , \theta, \phi ) &=& s_\sigma\, \sigma (-\tau , \pi - \theta ,\phi + \pi)\label{par5} \:,\\
k_{ab}(\tau , \theta, \phi ) &=& s_k \, k_{ab} (-\tau , \pi - \theta ,\phi + \pi)\, .\label{par52}
\eea
and choose the spatial boundary counter-term such that $\mathcal B_{S_\pm}$ is finite or zero. Here, $s_\sigma$ and $s_k$ are two signs which define the phase space with parity conditions. The Hamiltonian parity conditions imposed in \cite{Regge:1974zd} amount to $s_\sigma = s_k =+1$\footnote{ A necessary and sufficient condition to allow the Schwarzschild black hole is $s_\s = +1$. The choice $s_k = +1$ is necessary in order to disallow parity-odd supertranslations and therefore non-trivial supertranslation charges, as can be shown from the explicit expression \eqref{Qsupertransl} below. Together, these parity conditions are equivalent to the parity conditions of Hamiltonian fields imposed in \cite{Regge:1974zd}, see dictionary in Appendix \ref{app:BC}.}.

Let us now show that there is another mechanism for canceling the logarithmic divergence without assuming that $\s$ or $k_{ab}$ obey parity conditions. We simply propose to choose the spatial boundary counter-term to be 
\bea
\mathcal S_{\mathcal H}  = \mathcal S^{part}_{\mathcal H} +\frac{\log\Lambda}{4 \pi G} \Big( S^{(\s)} + S^{(k)}\Big)   \, ,\label{action}
\eea
where $S^{(\s)}$ and $S^{(k)}$ are the actions
\bea
S^{(\s)} &=&     \int_{\mathcal{H}} d^3 x  \sqrt{-h^{(0)}} \: \Big( -\frac{1}{2}  \: \DD_a \s \DD^a \s +\frac{3}{2}\s^2 \Big)  \label{Ss},\\
S^{(k)} &=&\frac{1}{4} \int_{\mathcal{H}} d^3 x \sqrt{-h^{(0)}} \: \Big( - \frac{1}{4} \DD_a k_{bc}\DD^ak^{bc}+\frac{1}{2}\DD_a k_{bc}\DD^b k^{ac}+\frac{1}{4}\p_a k \p^a k\nn\\
&& -\frac{1}{2}\p_a k \DD_b k^{ab}+(k_{ab}k^{ab}-\frac{1}{2}k^2)-\frac{M^2}{4}(k_{ab}k^{ab}-k^2)\Big),
\label{Sk}
\eea
for the scalar $\s$ and the traceless, divergence-free tensor $k_{ab}$ defined on the hyperboloid. Here, 
\bea
M^2 = 1.\label{M2}
\eea

Also, $\mathcal S^{part}_{\mathcal H}$ is a counter-term action built algebraically from the boundary fields (and that therefore contain no logarithmic divergence in $\mathcal B_{S_\pm}$) and that is chosen to cancel the boundary terms at spatial infinity, e.g. the Mann-Marolf counterterm \cite{Mann:2005yr}. We will discuss that counter-term action below. 
%Finally, the action $S_{(0)}$ is introduced for completeness. It can be any finite action defined on the boundary hyperboloid $\mathcal H$ which is zero on-shell, e.g. a combination of the actions $S^{(\s)}$ and $S^{(k)}$.

The variations of the actions $S^{(\s)}$ and $S^{(k)}$ reduce on-shell to a boundary term at $S_\pm$ because the equations of motion for $S^{(\s)}$ and $S^{(k)}$ are precisely the equations obeyed by $\s$ and $k_{ab}$ obtained from Einstein's equations, see Section \ref{sec:EOM}. Now, the sum of logarithmic divergences of the bulk term and boundary counter-term at $S_\pm$ precisely cancel for the choice of coefficients in \eqref{action} and the choice of boundary terms in the actions \eqref{Ss}-\eqref{Sk}. 

The action \eqref{Sk} with $M^2=0$ is the Fierz-Pauli action for a massless spin 2 field in de Sitter spacetime. For $M^2 = 1$, it is the three-dimensional version of the action discussed earlier in a different context by Deser and Nepomechie \cite{Deser:1983mm} and Higuchi \cite{Higuchi:1989gz}. For the particular value of the mass \eqref{M2}, the action is invariant under the supertranslation transformation law \eqref{actsuper}, as it should by consistency with the bulk equations of motion. The divergence of the equations of motion implies 
$
\DD^a k_{ab} = \p_b k
$. Also, the trace $k$ of $k_{ab}$ is unconstrained by the equations of motion. The action is therefore compatible with our boundary condition $k=\DD^b k_{ab}=0$. 

The action \eqref{action} explicitly breaks translation, log translation and supertranslation invariance but does not break Lorentz invariance. The presence of logarithmic counter-terms is reminiscent of the Weyl anomaly \cite{Capper:1974aa,Deser:1976aa} in the holographic renormalization of anti-de Sitter spacetimes in odd spacetime dimensions \cite{Henningson:1998gx}. We will refer to the action
\bea
\mathcal A &=& \frac{1}{4\pi G} (S^{(\s)}  +S^{(k)}  ),\label{anoaction}
\eea
as the (super/log-)translation anomaly. The anomaly is invariant under all symmetries that are broken. Indeed, translations do not act on the fields $\sigma$ and $k_{ab}$. Logarithmic translations and supertranslations act as \eqref{actlog} and \eqref{actsuper} but the anomaly is invariant up to boundary terms at timelike boundaries (that we neglect for this argument). Therefore, the Noether charges of the anomaly associated with the (super/log)-translation symmetries represent the algebra of (super/log)-translations. The Wess-Zumino consistency conditions \cite{Wess:1971aa} are therefore obeyed.
%, see e.g. \cite{Skenderis:2000in} for the analogous statement for the Weyl anomaly in anti-de Sitter spacetimes.
Even though the anomaly is zero on-shell for all metrics obeying the boundary conditions, it affects the dynamics mainly because its symplectic structure is non-zero on-shell, as we will discuss in Section \ref{sec:symplectic}. Since no holographic model for asymptotically flat spacetimes is known, we cannot unfortunately try to match the flat spacetime anomaly to a QFT model.

When parity conditions hold, no logarithmic divergent term appears in the variation of the Einstein-Hilbert action. There is therefore no requirement to  add an anomaly term in the Einstein-Hilbert action. If one insists in defining an unique action principle whether or not parity conditions hold, the phase space of asymptotically flat gravity where parity conditions do hold also has a non-vanishing anomaly in the action. The variation of the anomaly does not lead to any logarithmic divergent boundary term because of parity conditions, which is consistent with the fact that there is no divergent term to cancel. The anomaly therefore does not contribute to the symplectic structure and the conserved charges, as we will detail below. More generally, the anomaly does not participate in any on-shell dynamics. For all purposes in classical gravity, one can therefore ignore the anomaly. The anomaly might however have a role in a quantum path integral involving off-shell configurations. Since this work is purely classical, we will simply discard the anomaly when the phase space is restricted with parity conditions. 

Let us now discuss the remaining spatial counterterm $\mathcal S^{part}_{\mathcal H}$. Following earlier work \cite{BY,Kraus:1999di, Mann:1999pc, deHaro:2000wj,Astefanesei}, it was proposed in \cite{Mann:2005yr} to define $\mathcal S^{part}_{\mathcal H}$ as 
\begin{equation}
 \mathcal S^{part}_{\mathcal H}= \frac{1}{8\pi G} \int_{\rho = \Lambda} d^3 x \sqrt{-h} \,(K - \hat K) ,
\end{equation}
where $K$ is the trace of the extrinsic curvature, $\hat K \equiv h^{ab} \hat K_{ab}$ and $\hat K_{ab}$ is defined implicitly by
\begin{equation}
\label{Khat}
{\cal R}_{ab} = \hat K_{ab} \hat K - \hat K_a{}^{c} \hat K_{cb},
\end{equation}
where ${\cal R}_{ab} $ is the Ricci tensor of the boundary metric $h_{ab}$. This equation being quadratic in $\hat K_{ab}$, it admits more than one solution for $\hat{K}_{ab}$. The prescription of \cite{Mann:2005yr} consists of choosing the solution that asymptotes to the extrinsic curvature of the boundary of Minkowski space as $\Lambda$ is taken to infinity. It can then be shown that the variation of the action at the spatial hyperbolic boundary is equal on-shell to
\begin{eqnarray}
\delta S |_{\rho = \Lambda} = \frac{1}{16 \pi G} \int_{\mathcal{H}} d^3 x \sqrt{-h^{(0)}}\:  E^{(1)\: ab} \delta k_{ab}  ,
\end{eqnarray}
where $E_{ab}^{(1)} = -\s_{ab} - \s h_{ab}^{(0)}$ is the first order electric part of the Weyl tensor. The variation of the anomaly action  \eqref{anoaction} is zero on-shell (up to the crucial boundary term $\mathcal B_{S_\pm}$ treated earlier) and therefore does not contribute on the boundary hyperboloid. Here, Einstein's equations imply that $h_{ab}^{(0)}$ is locally the metric on the hyperboloid and therefore we set $\delta h_{ab}^{(0)} = 0$ by fixing the boundary metric to be the unit hyperboloid.

Now, let us observe that the condition $\delta k_{ab} = 0$ imposed in \cite{Mann:2005yr} can be relaxed. Using integrations by parts, the boundary conditions \eqref{condk} imply that
\begin{eqnarray}
\delta S |_{\rho = \Lambda} = \frac{1}{8\pi G} \int_{\mathcal{H}} d^3 x \sqrt{-h^{(0)}} \s \delta \bar k = 0\, .
\end{eqnarray}
As already mentioned, we expect that one can constraint the initial and final states and add appropriate boundary terms at past and future times in order to cancel the boundary terms at $\Sigma_\pm$ in the variation of the action $\delta S$. Therefore, up to that assumption, the action \eqref{covaction} is a valid variational principle for all $k_{ab}$ obeying \eqref{condk} without necessarily imposing parity conditions. Note that the trace $k$ is conjugated to the mass aspect ratio $\s$ in the variational principle, which is reminiscent of the AdS variational principle where the boundary metric is conjugated to the stress-energy tensor \cite{de Haro:2000xn,Skenderis:2000in}\footnote{We thank K. Skenderis for pointing this out.}.

If one insists in imposing parity conditions on $\s$  and $k_{ab}$, the covariant phase space defined by our boundary conditions is still more general than the one considered in \cite{ABR,Ashtekar:2008jw} since here we do not impose $k_{ab}=0$ but only require that $\DD^a k_{ab} = k_a^a=0$, which implies that the first order part of the magnetic Weyl tensor $B^{(1)}_{ab}= \frac{1}{2}\epsilon_a^{\;\; c d} \DD_c k_{d b} $ can be fluctuating and non-vanishing\footnote{Note that the four lowest harmonics in $B_{ab}^{(1)}$ are zero because we impose that $k_{ab}$ is regular, see Lemma 1 in Appendix \ref{app:Nut}.}. In our phase space with parity conditions, the parity even supertranslations fulfilling \eqref{super} are allowed while logarithmic translations are forbidden. As we will see herebelow in Section \ref{LagCharges}, the conserved charges associated with those allowed supertranslations are vanishing while the Poincar\'e generators are non-vanishing. Therefore, the phase space defined by our boundary conditions \eqref{metric1} supplemented by parity conditions is a consistent phase space where the asymptotic symmetry group is the Poincar\'e group.
 The covariant phase space defined by \eqref{metric1} with parity conditions is a subset of the phase space defined in 3+1 formalism by Regge and Teitelboim in \cite{Regge:1974zd} where angular supertranslations have been fixed by requiring $g_{\rho a}= o(\rho^{-2})$.

In the following, we will derive the dynamics of the phase space defined from our boundary conditions from the action principle. Since we regularized all divergences the symplectic structure and conserved charges will be finite when all boundary contributions are taken into account.

\subsection{Equations of motion}
\label{sec:EOM}

Using the 3+1 split of Einstein's equations along the hyperbolic radial coordinate $\rho$ and using our expression for the metric (\ref{metric1}), we can expand the equations up to second order in the $\rho$ expansion. This analysis is performed in Appendix \ref{appEOM}. By convention, we refer to the equations involving only $h_{ab}^{(0)}$ as the zeroth order equations, the ones linear in $k_{ab}$, $\s$ as the first order equations, and the ones linear in $i_{ab}$ and $h_{ab}^{(2)}$ as second order equations. The zeroth and first order equations were already obtained in full generality in \cite{BS}. As already said before, the zeroth order equations imply that $h_{ab}^{(0)}$ is locally the metric of the unit hyperboloid. At first order, setting $k_{ab}$ to be traceless and divergenceless, we obtain
\bea
(\square + 3) \sigma = 0 \; ,\qquad (\square -3)k_{ab}  =0 \; .\label{EOM1a}
\eea

One can rewrite elegantly those first order equations in terms of the  first order electric and magnetic parts of the Weyl tensor
\bea
E^{(1)}_{ab} &=& -\DD_a \DD_b \sigma - h^{(0)}_{ab} \sigma \, ,\qquad
B^{(1)}_{ab} =  \frac{1}{2}\epsilon_a^{\;\; c d} \DD_c k_{d b} \, .\label{dBk}
\eea
By construction, these two tensors enjoy the following properties
\bea
h^{(0)ab} B^{(1)}_{ab} = 0\; , \qquad  E^{(1)}_{[ab]}  = 0\; , \qquad  \DD_{[c} E^{(1)}_{a] b} = 0.
\eea
The boundary conditions \eqref{condk} imply that $B^{(1)}_{[ab]}  = 0$.
The equations of motion at first order \eqref{EOM1a} are then equivalent to
\bea\label{EOM1bis}
h^{(0)ab} E^{(1)}_{ab} = 0\; ,\qquad  \DD_{[c} B^{(1)}_{a] b} = 0.
\eea

At second order, the equations derived in the literature do not include $k_{ab}$ and $i_{ab}$. A complete derivation is provided in Appendix \ref{appEOM} and the resulting equations are given from  (\ref{eqh2a}) to (\ref{eqh2d}). Instead of giving their explicit cumbersome expressions, and following the works presented in \cite{B} and \cite{Compere:2011db}, these equations can be reformulated in an interesting and compact way by introducing a pair of conjugate symmetric, divergence-free and traceless (SDT) tensors $V_{ab}$ and $W_{ab}$. They are defined as
\bea
V_{ab} &\equiv& - h_{ab}^{(2)}+\frac{1}{2} i_{ab}+ Q_{ab}^V \; , \label{defnlV} \\
W_{ab} &\equiv& \eps_a^{\;\, cd}D_c \left( h^{(2)}_{d b }-\frac{1}{2} i_{db} + Q_{db}^W \right) \; , \label{defnlW}
\eea
where $Q_{ab}^V$ and $Q_{ab}^W$ are non-linear terms defined in appendix in \eqref{defQV}-\eqref{defQW}. The tensors are conjugate in the sense that they obey the following duality properties
\bea
W_{ab} = - (\curl V)_{ab}  +  (\curl \kappa)_{(ab)} ,\qquad V_{ab} = (\curl W)_{ab} -2 i_{ab} \, ,\label{nl03}
\eea
where
\beqs
\kappa_{ab} &=& \kappa^{[\s,\s,I]}_{ab} %- \kappa^{[2,\s,\s,II]}_{ab}
+ 4 \kappa^{[\s,k, I]}_{ab} + Y^{(2)}_{ab}+ \kappa^{[k,k,I]}_{ab}  \label{defkappa}
%+ \kappa^{[2,k,k,III]}_{ab} \: ,
\eeqs
is a sum of symmetric and divergence-free (SD) tensors. All SD tensors built out of quadratic terms in $\s$, $k_{ab}$ and their derivatives can be completely classified. This classification is performed in Appendix \ref{appSD}. The particular tensors appearing in $\kappa_{ab}$ \eqref{defkappa} and some of their important properties are described at length in Appendix \ref{appSD}.

Using the SDT tensor $W_{ab}$,  it is shown in Appendix \ref{equivsys} that the equations of motion can be written as
\bea
W^a_a &=& \DD^b W_{ab}=0, \nonumber \\
(\square - 2)W_{ab} &=& \curl(2 i + \kappa )_{(ab)} ,\label{eqWf}\\
i^a_a &=& \DD^b i_{ab} = 0 ,\nonumber\\
(\square -2)i_{ab} &=& 0.\label{eqi}
\eea
Using the curl operator and the definition $j_{ab} \equiv -(\text{curl} \: i)_{ab}$, one can derive an equivalent form of those equations in terms of $V_{ab}$ as
\bea
V^a_a &=& \DD^b V_{ab}=0 ,\nonumber \\
(\square - 2)V_{ab} &=& \curl(-2j + \curl(\kappa ))_{ab},\label{eqVf} \\
j^a_a &=& \DD^b j_{ab} = 0, \nonumber\\
(\square -2)j_{ab} &=& 0.\label{eqj}
\eea
Since the curl of the latter set of equations lead to \eqref{eqWf}-\eqref{eqi}, the two sets are equivalent. Once the set of equations \eqref{eqWf}-\eqref{eqi} is solved, one can reconstruct $h_{ab}^{(2)}$ from the definitions \eqref{defnlV} or \eqref{defnlW}. Einstein's equations at second order are therefore equivalent to either set of the above systems of equations.

\subsection{Linearization stability constraints}
\label{sec:LS}

For any SDT tensor $T_{ab}$ and any rotation or boost vector $\xi_{(0)}^a$, such that $\DD_{(a}\xi_{(0) b)} = 0$ and $(\square +2)\xi_{(0)}^a = 0$, we get
\bea
(\Box - 2)T_{ab}\: \xi_{(0)}^a = 2 \: \DD^a \left( \xi_{(0)}^c \DD_{[a}T_{b]c}+T_{c[a}\DD_{b]} \xi_{(0)}^c \right).\label{eq:86t}
\eea
Therefore, the following integral
\bea
\int_S d^2S \: (\Box - 2)T_{ab}\: \xi_{(0)}^a \:  n^b = 0
\eea
vanishes as the r.h.s. of \eqref{eq:86t} reduces to a total divergence on the two-sphere. As pointed out in \cite{B}, necessary conditions for the equations of motion  \eqref{eqWf} to admit solutions are that the r.h.s. of \eqref{eqWf} contracted with $\xi_{(0)}^a$ and integrated on the sphere vanishes
\bea
\int_S d^2S \: (\Box - 2) W_{ab}\: \xi_{(0)}^a \:  n^b= \int_S d^2S \: \curl(2 i + \kappa )_{(ab)} \: \xi_{(0)}^a n^b = 0 \: .
\eea
Such conditions were referred to as integrability conditions in \cite{B}. They can be recognized as linearization stability constraints \cite{Moncrief:1975aa,Moncrief:1976un}.

Using the fact that charges constructed with the curl of an SD tensor (such as $\kappa_{ab}$ or $i_{ab}$) can be mapped to charges associated with the SD tensor itself, as proved in the appendix B of \cite{Compere:2011db}, these conditions are equivalent to
\bea
\int_S d^2S \, i_{ab} \: \xi_{(0)}^a n^b = -\frac{1}{2} \int_S d^2S \, \kappa_{ab} \: \xi_{(0)}^a n^b \, .\label{condI2}
\eea
Now, using the properties of $\kappa_{ab}$ detailed in Appendix \ref{appSD}, it turns out that there exists a unique  equivalence class of tensor structures quadratic in $\s$, respectively in $k_{ab}$, which might evaluate to a non-zero result in the r.h.s of \eqref{condI2}. For the mixed $(\s,k)$ SD tensors, we show that none of them could contribute.

Remarkably, the two equivalence classes of charges quadratic in either $\s$ or $k_{ab}$ can be recognized as the Noether charges associated with the two free actions \eqref{Ss}-\eqref{Sk}. This provides an independent argument that the actions  \eqref{Ss} and \eqref{Sk} play an important role in the dynamics of asymptotically flat spacetimes. More precisely, using the Noether charges of the actions \eqref{Ss}-\eqref{Sk} as representatives of the two classes of conserved charges that one can build out of quadratic terms in $\s$ and $k_{ab}$, we can rewrite the linearization stability constraints as
\bea
\int_S d^2S\,  i_{ab} \: \xi_{(0)}^a n^b =  2\:  \int_S d^2S  \, \Big(  T^{(\s)}_{ab} + T^{(k)}_{ab} \Big) \xi_{(0)}^a n^b , \label{intFF}
\eea
where
\bea
T^{(\s)}_{ab} &\equiv& - \frac{2}{\sqrt{-h^{(0)}}} \frac{\delta L^{(\s)}}{\delta h^{(0)\: ab}} = -\frac{1}{4} \kappa^{[\s,\s,I]}_{ab} +\frac{1}{2} \kappa^{[\s,\s,II]}_{ab} , \label{Ts} \\
T^{(k)}_{ab} &\equiv& -  \frac{2}{\sqrt{-h^{(0)}}} \frac{\delta L^{(k)}}{\delta h^{(0) \: ab}}
%&=& 4 B_{(1)}^{ac}B_{(1)c}^a - h_{(0)}^{ab} B_{(1)cd}B_{(1)}^{cd} + \eps^{cd(a} ( \DD_{c} k_e^{\:\: b)} B^{(1)\:e}_{d}+ \DD^{b)}B^{(1)\:e}_{c} k_{de} ) \nonumber \\
= \frac{1}{2} \kappa^{[k,k,II]}_{ab} -\frac{1}{8} Y^{(2)}_{ab} , \label{Tk}
\eea
are the stress-tensors of the actions $S^{(\s)}$ and $S^{(k)}$ given in \eqref{Ss}-\eqref{Sk}. The SD tensors appearing on the r.h.s of \eqref{Ts} and \eqref{Tk} are defined in Appendix \ref{sec:SD}.

\subsection{Symplectic structure}
\label{sec:symplectic}

In this section we first review the standard covariant phase space symplectic structure \cite{AshtekarU,CrnkovicWitten,Lee:1990nz}. We then fix the well-known ambiguity \cite{Wald:1993nt,Iyer:1994ys} in its definition using the boundary terms in the action principle \eqref{action}. Effectively, we will fix the boundary terms in the symplectic structure in such a way that the logarithmic divergences, present when parity conditions are not imposed, cancel. Our procedure then amounts to the prescription already proposed in \cite{CompereMarolf}. The final symplectic structure that we define has the form
\bea
\Omega[\delta_1 g,\delta_2 g] = \Omega_{\text{bulk}}[\delta_1 g,\delta_2 g] +  \Omega_{\text{c.t.}}[\delta_1 \s,\delta_1 k ; \delta_2 \s,  \delta_2 k]  , \label{OmegatotL}
\eea
where $\Omega_{bulk}$ is the standard bulk symplectic structure and $\Omega_{c.t.}$ is precisely the boundary symplectic structure that can be derived from the boundary action in \eqref{action}.

The symplectic structure is a phase space 2-form defined as an integral over a Cauchy slice of a spacetime 3-form that we refer to as the integrand for the symplectic structure. The Ashtekar-Crnkovic-Witten-Lee-Wald integrand \cite{AshtekarU,CrnkovicWitten,Lee:1990nz} for the covariant phase space symplectic structure is given by
\bea
\omega [\delta_1 g ,\delta_2 g ] &=& \frac{1}{32\pi G}(d^{3}x)_{\mu}\sqrt{-g } \Big(
\delta_1 g^{\alpha\beta}\nabla^\mu \delta_2 g_{\alpha \beta}+\delta_1 g \nabla^\alpha \delta_2 g^\mu_{\alpha}+\delta_1 g^\mu_{\alpha}\nabla^\alpha \delta_2 g  \nn \\
&&  - \delta_1 g \nabla^\mu \delta_2 g - 2 \delta_1 g_{\alpha\beta}\nabla^\alpha \delta_2 g^{\mu\beta}
%- \delta_1 g^{\mu\alpha}\nabla^\beta \delta_2 g_{\alpha \beta}
- (1 \leftrightarrow 2)
\Big),
\eea
where $\delta_1 g_{\mu\nu}$, $\delta_2 g_{\mu\nu}$ are perturbations around a general asymptotically flat spacetime $g_{\mu\nu}$.
This integrand is obtained by varying a second time the boundary term $\Theta [\delta g]$ obtained after a variation of the Einstein-Hilbert Lagrangian as $\omega[\delta_1 g,\delta_2 g] = \delta_1 \Theta[\delta_2 g] - \delta_2 \Theta[\delta_1 g]$. The integrand for the symplectic structure,  obtained in  \cite{Barnich:2007bf} by acting with a contracting homotopy on Einstein's equations contracted with $\delta g_{\mu\nu}$, is given instead by
\bea
W [\delta_1 g ,\delta_2 g ] &=& \frac{1}{32 \pi G} (d^3x)_{\mu} \: \sqrt{-g}\: P^{\mu\nu\alpha\beta\gamma\delta} \Big( \delta_1 g_{\alpha \beta} \n_\nu \delta_2 g_{\gamma \delta} -\delta_2 g_{\alpha \beta} \n_\nu \delta_2 g_{\gamma \delta} \Big) \nonumber \\
&=& \frac{1}{32\pi G}(d^{3}x)_{\mu}\sqrt{-g } \Big(
\delta_1 g^{\alpha\beta}\nabla^\mu \delta_2 g_{\alpha \beta}+\delta_1 g \nabla^\alpha \delta_2 g^\mu_{\alpha}+\delta_1 g^\mu_{\alpha}\nabla^\alpha \delta_2 g \nn \\
&&  - \delta_1 g \nabla^\mu \delta_2 g - \delta_1 g_{\alpha\beta}\nabla^\alpha \delta_2 g^{\mu\beta} - \delta_1 g^{\mu\alpha}\nabla^\beta \delta_2 g_{\alpha \beta}
- (1 \leftrightarrow 2)
\Big),
\eea
where
\beqs
P^{\mu\nu\alpha \beta \gamma \delta}&=&g^{\mu\nu} g^{\gamma(\alpha} g^{\beta)\delta} + g^{\mu(\gamma} g^{\delta)\nu} g^{\alpha \beta}+ g^{\mu(\alpha} g^{\beta)\nu} g^{\gamma \delta} \nonumber \\
 &&-g^{\mu\nu} g^{\alpha \beta} g^{\gamma \delta}-g^{\mu (\gamma} g^{\delta)(\alpha} g^{\beta)\nu}-g^{\mu(\alpha} g^{\beta)(\gamma} g^{\delta)\nu}.
 \eeqs

The integrands of the two symplectic structures defined above differ by the boundary term
\bea
W [\delta_1 g ,\delta_2 g ] - \omega [\delta_1 g ,\delta_2 g ]  = \frac{1}{32\pi G}(d^{3}x)_{\mu}\sqrt{-g } \nabla_\nu \Big( \delta_1 g^\nu_{\; \beta} \delta_2 g^{\mu\beta} - (\mu \rightarrow \nu)\Big),
\eea
which vanishes on constant $\rho$ and $\tau$ surfaces when the metric and perturbations are expanded in the Beig-Schmidt expansion with
\bea
g_{\rho a}=0,\qquad \delta g_{\rho a} = 0\, .
\eea
We can therefore use interchangeably $\omega$ and $W$ in what follows. The integrand for the symplectic structure derived from Einstein's equations contracted with the Lie derivative of the metric is a boundary term
\bea
W[\delta g , \mathcal L_\xi g ] % &\equiv&  \frac{1}{32 \pi G} (d^3x)_{\mu}\: \sqrt{-g} \: P^{\mu\nu\alpha\beta\gamma\delta} \Big( \delta g_{\alpha \beta} \n_\nu (\n_{\gamma} \xi_{\delta} +\n_{\delta} \xi_{\gamma}) - (\n_{\alpha} \xi_{\beta} +\n_{\beta} \xi_{\alpha}) \n_\nu \delta g_{\gamma \delta}\Big) \nonumber \\
&=& d k_\xi [ \delta g ; g ],\label{propsym}
\eea
where the last equality has to be understood up to Einstein's equations of motion for the metric and the linearized perturbations. The boundary term $k_\xi [ \delta g ; g ]$ is exactly given by the Abbott-Deser expression \cite{Abbott:1981ff} for the surface charge for a linear perturbation $\delta g_{\mu\nu}$ around a solution $g_{\mu\nu}$
\bea
k_\xi [\delta g ; g ] &=&  \frac{2}{3} \: \frac{1}{16\pi G} \: \sqrt{-g}\: (d^2 x)_{\mu\alpha} P^{\mu\nu\alpha\beta\gamma\delta} \Big( 2 \xi_{\nu} \n_{\beta} \delta g_{\gamma \delta} -\delta g_{\gamma \delta} \n_{\beta} \xi_{\nu} \Big) \nonumber \\
&=& \frac{1}{16\pi G}(d^{2}x)_{\mu\nu}\sqrt{-g } \Big( \xi^\nu (\D^\mu \delta g - \D_\sigma \delta g^{\s \mu}) +\xi_\sigma \D^\nu \delta g^{\s \mu} \nonumber \\
&&+\frac{1}{2} \delta g \D^\nu \xi^\mu +\frac{1}{2}\delta g^{\mu \sigma}\D_\sigma \xi^\nu +\frac{1}{2} \delta g^{\nu \sigma}\D^\mu \xi_\sigma - (\mu \leftrightarrow \nu)
\Big) .
\label{AD}
\eea

Let us now discuss conservation and finiteness of the symplectic structure. On the one hand, using the expansion \eqref{BC2} at first order, the symplectic structure integrand evaluated on a hypersurface $\rho =$ constant gives
\bea
W [\delta_1 g ,\delta_2 g ]|_{fixed\; \rho}  = o(\rho^{0}),
\eea
which shows that the symplectic structure is conserved. On the other hand, the integrand for the symplectic structure evaluated on a Cauchy slice $\Sigma$ asymptotic to a constant $\tau$ hypersurface reads
\bea
W [\delta_1 g ,\delta_2 g ]|_\Sigma &=& \frac{\rho^{-1}}{4\pi G} (d^3 x)_a \sqrt{-h^{(0)} } \Big( \delta_1 \sigma \DD^a \delta_2 \sigma  -\frac{1}{4} \epsilon^{a c e }\:  \delta_1 k_c^{\:\:d} \: \delta_2 \:  B^{(1)}_{ed} - (1 \leftrightarrow 2) \Big) + o(\rho^{-1}) , \label{Weval} \nonumber \\
\eea
which is consistent with the earlier expression \eqref{varS2}. The bulk symplectic structure
\bea
\Omega_{bulk}[\delta_1 g,\delta_2 g] \equiv \int_\Sigma W [\delta_1 g ,\delta_2 g ]
\eea
is therefore logarithmic divergent for generic $\sigma$ and $k_{ab}$, in accordance with the variational principle being ill-defined on future and past boundaries $\Sigma_\pm$ when parity conditions are not imposed.

We have seen in Section \ref{sec:action} that there is a mechanism to cancel the logarithmic divergence in the action without imposing parity conditions. The logarithmic divergences are canceled once the Einstein-Hilbert action is supplemented by a logarithmic counter-term, as in \eqref{action}. Now, even though the logarithmic counter-term action is zero on-shell (up to boundary terms), it has a non-vanishing boundary contribution to the symplectic structure. Indeed, the free actions for $\s$ and $k_{ab}$ introduced in \eqref{Ss}-\eqref{Sk} are zero on-shell but their symplectic structures  (defined with the same conventions as in the bulk) are the Klein-Gordon norm and the symplectic norm between two traceless transverse fields given by the integral of
\bea
\omega_{(\s)}[\delta_1 \s , \delta_2 \s] &=& (d^2 x)_a \sqrt{-h^{(0)}} \Big( \delta_1 \s \p^a \delta_2 \s  - (1 \leftrightarrow 2) \Big),\label{omegas} \\
\omega_{(k)}[\delta_1 k , \delta_2 k] &=& (d^2x)_a \sqrt{-h^{(0)}}  \Big( \frac{1}{4}\eps^{adc}\delta_1 B^{(1)b}_c  \delta_2 k_{db}- (1 \leftrightarrow 2) \Big) \, ,
\label{omegak}
\eea
over the sphere which is generally non-vanishing\footnote{The counter-term $\int_{\rho=\Lambda}d^3 x \sqrt{-h}(K-\hat K)$ has not the form of an off-shell action for the boundary fields and therefore it does not define a boundary symplectic structure.}. The total symplectic structure is then defined as announced in \eqref{OmegatotL} with
\bea
\Omega_{c.t.}[\delta_1 \s,\delta_1 k , \delta_2 \s,  \delta_2 k]  =  \frac{\log\Lambda}{4\pi G} \int_S \Big(
 \omega^{(\s)}[\delta_1 \s , \delta_2 \s] +  \omega^{(k)}[\delta_1 k , \delta_2 k] \Big)  \, . \label{Omegact}
\eea
The resulting prescription for fixing the boundary terms in the symplectic structure left unfixed in \cite{Lee:1990nz,Wald:1993nt,Iyer:1994ys} amounts to the prescription argued in \cite{CompereMarolf} to fix the boundary terms in the symplectic structure using the symplectic structure of the boundary terms of the action. We see that we can now justify that prescription by the existence of a variational principle when past and future boundaries are taken into account.

\subsection{Covariant phase space charges}
\label{LagCharges}

In this section, we compute the covariant phase space charges, paying particular attention to the contributions coming from the boundary symplectic structure \eqref{Omegact}. Since the symplectic structure is finite, the conserved charges are also finite. Along the way, we review how the infinitesimal charge coming from the bulk symplectic structure (which is not finite in itself when parity conditions are not imposed) can be expressed equivalently in the form derived by Abbott-Deser \cite{Abbott:1981ff}, Iyer-Wald \cite{Wald:1993nt,Iyer:1994ys} or Barnich-Brandt \cite{Barnich:2001jy}. After some general considerations, we respectively consider the conserved charges associated with translations, supertranslations, Lorentz charges and logarithmic translations. When parity conditions are imposed, in addition to our boundary conditions, the resulting expressions for the charges reduce to the ones derived in previous works on asymptotically flat spacetimes in Lagrangian framework \cite{Geroch:1977aa,Ashtekar:1978zz,AshRev,Abbott:1981ff,Ashtekar:1991vb}. Indeed, as we have already stated, the only difference with those works, when parity conditions are imposed, is that we have allowed for temporal and radial  supertranslations whose associated charges are zero.

The covariant phase space infinitesimal charges associated with a diffeomorphism tangent to the phase space are defined from the symplectic structure as
\bea
\slash \hspace{-5pt} \delta Q_\xi [g] = \Omega [\delta g , \mathcal L_\xi g  ] \, .
\eea
Here, we use the symbol $\slash\hspace{-5pt} \delta Q_\xi[g]$ to remind the reader that the infinitesimal charge between the solution $g$ and $g+\delta g$ can be considered as a one-form in field space which is not necessarily exact. When $\slash\hspace{-5pt} \delta Q_\xi[g]$ is an exact form in phase space, as it will turn out to be the case for each diffeomorphism that we consider, the charges can be defined. We denote the integrated charge as $Q_\xi[g ; \bar g]$ (we have $\delta Q_\xi[g ; \bar g] = \slash\hspace{-5pt} \delta Q_\xi[g]$) and fix the integration constant so that $Q_\xi[\bar g ; \bar g] = 0$ for Minkowski spacetime $\bar g$.

Using the definition \eqref{OmegatotL} and the property \eqref{propsym} of the bulk symplectic structure, the charge one-form $\slash\hspace{-5pt} \delta Q_\xi[g]$ can be written as a surface integral
\bea
\slash\hspace{-5pt} \delta Q_\xi[g] = \int_S k_\xi [\delta g ; g ] + \frac{\log\Lambda}{4\pi G} \int_S  \Big(
 \omega^{(\s)}[\delta \s , \delta_\xi \s] + \omega^{(k)}[\delta k , \delta_\xi k] \Big), \label{totkk}
\eea
evaluated on the sphere $S$ at constant time $t$ and at $\rho = \Lambda$. Here, $\delta_\xi \s$, $\delta_\xi k_{ab}$ are variations of the first order fields induced by the Lie derivative of the metric along $\xi$. The Abbott-Deser expression can be rewritten in the alternative form
\bea
\hspace{-6pt}k_\xi [\delta g ; g ] &=& - \delta k^K_\xi +  \frac{1}{8\pi G}(d^{2}x)_{\mu\nu}\sqrt{-g } \Big( \xi^\nu (\D^\mu \delta g - \D_\sigma \delta g^{\s \mu})+\D^\mu \delta \xi^\nu \Big) - E[\mathcal L_\xi g, \delta g]\label{kk1},
\eea
where
\bea
k^K_\xi[g] =  \frac{1}{16\pi G}(d^{2}x)_{\mu\nu}\sqrt{-g } \Big( \D^\mu \xi^\nu - \D^\nu \xi^\mu \Big)\, ,
\eea
is the Komar term and
\bea
E[\mathcal L_\xi g, \delta g] =   \frac{1}{16\pi G}(d^{2}x)_{\mu\nu}\sqrt{-g } \Big( \frac{1}{2} g^{\mu \alpha} \delta g_{\alpha \beta }g^{\beta \gamma}\mathcal L_{\xi}g_{\gamma \sigma}g^{\sigma\nu} - (\mu \leftrightarrow \nu)  \Big)\, , \label{kk3}
\eea
is a term linear in the Killing equations that might not vanish in general for asymptotic symmetries.  Here $\delta$ acts on the metric and on the asymptotic Killing vector, as  $\xi(g)$ might depend on the metric. Remark that in \eqref{kk1}, the first term is the exact variation of the Komar term and the third term is zero when evaluated on constant $\rho$ and $\tau$ surfaces since
\bea
g_{\rho a}=0,\qquad \delta g_{\rho a} = 0,\qquad \mathcal L_\xi g_{\rho a} = 0.
\eea
Therefore the bulk surface charge one-form \eqref{kk1} is given by the generalization of the Iyer-Wald expression when the asymptotic Killing vector is allowed to depend on the metric
\bea
k_\xi [\delta g ; g ] = - \delta k^K_\xi[g]  +  \frac{1}{8\pi G} (d^{2}x)_{\mu\nu}\sqrt{-g } \Big( \xi^\nu (\D^\mu \delta g - \D_\sigma \delta g^{\s \mu})+\D^\mu \delta \xi^\nu \Big) .\label{kk2}
\eea
The explicit evaluation of the total charge one-form \eqref{totkk} for each asymptotic Killing vector (translations, supertranslations, rotations, boosts and logarithmic translations) is a straightforward exercise that we do not reproduce here in details. In the following, we simply present the key steps in the evaluation of the charges and state the final results.

\vspace{3pt} \noindent {\bf Translations}\vspace{3pt}

For translations,  it is relatively simple to see that
\bea
k_\xi [\delta g ; g ] = k_\xi [\delta g ; \bar g ] ,\qquad  \delta_\xi \s = \delta_\xi k_{ab} = 0 .\label{aslin}
\eea
As a result, the boundary contributions to the charges vanish and the charges can be defined using linearized perturbations of asymptotic fields around Minkowski spacetime
\bea
\mathcal Q_\xi[g ; \bar g]  = \int_S k_\xi [h ; \bar g ] ,\qquad h_{\mu\nu} \equiv g_{\mu\nu}- \bar g_{\mu\nu}\, .
\eea
Effectively, one can therefore derive the charges associated with translations using the bulk linearized theory and obtain the well-known expression for the four momenta
\bea
Q_{(\mu)}[g ; \bar g ] = - \frac{1}{8 \pi G} \int_S d^2S E_{ab}^{(1)} n^a \DD^b \zeta_{(\mu)}\, ,\label{ch:tr}
\eea
where the four scalars $\zeta_{(\mu)}$, $\mu=0,1,2,3$ are the four solutions of $\DD_a \DD_b \zeta_{(\mu)} + h_{ab}^{(0)}\zeta_{(\mu)}= 0$.  The vector $\p/\p t$ in the 3+1 asymptotic spherical coordinate system $(t,r,\theta,\phi)$ corresponds to the translation $\zeta_{(0)} = -\sinh \tau$. One can check \cite{Mann:2008ay} that the charge $Q_{(0)}$ for the Kerr black hole of mass $m$ is $+m$ after identifying $\sigma = G m \cosh 2\tau \sech \tau$, as it should.

\vspace{3pt} \noindent {\bf Supertranslations}\vspace{3pt}

For supertranslations, we start by showing that
\bea
k_\xi [\delta g ; g ] = k_\xi [\delta g ; \bar g ] ,\qquad  \delta_\xi \s =0, \qquad  \delta_\xi k_{ab} = 2 (\omega_{ab}+h_{ab}^{(0)}\omega ) .
\eea
The bulk charges can again be obtained using the linearized theory. Now, it is important to note that the boundary counter-terms do not contribute to the supertranslation charges
\bea
\int_S  \omega_{(\s)} [ \delta \s ,\delta_\xi \s ] = 0,\qquad  \int_S  \omega_{(k)} [ \delta k ,\delta_\xi k ] =0\, .
\eea
The first equality is trivial. The second one can be proven after using $\delta_\xi B^{(1)}_{ab} = 0$,  $B^{(1)}_{ab} = - \hat \s_{ab} - h_{ab}^{(0)} \hat \s ,$ where $(\square + 3)\hat \s = 0$ as a consequence of Lemma 1 of \cite{Compere:2011db} (see also \cite{Ashtekar:1978zz}) and eventually the fact that $n_a \eps^{ace}\omega_c^d \hat \s_{ed} = n_a \DD_c \Big( \eps^{ace} (\frac{1}{2} \omega^d \hat \s_{ed} - \frac{1}{2}\hat \s^d \omega_{ed} - \omega \hat \s_e)  \Big)$ is a boundary term.

The supertranslation charges are therefore precisely the ones which can be obtained from the bulk linearized theory. They read as
\bea
Q_{(\omega)}[g ; \bar g ] =  \frac{1}{4\pi G}\int_S d^2 S \sqrt{-h^{(0)}} n_a \left( \s^a \omega - \s \omega^a    \right) .
\label{Qsupertransl}
\eea
As we restrict ourselves to supertranslations satisfying \eqref{super}, the charges are conserved since $\DD_a (  \s^a \omega -\s \omega^a   ) =  \square \s \omega-\s \square \omega  = 0$. Note that if we choose $\omega$ to be a translation, the charges indeed reduce to the earlier expression \eqref{ch:tr} upon using $2 \omega \s_b - 2 \omega_b \s   = -E^{(1)\:a}_b \omega_a - 2 \DD^a (\omega_{[a}\s_{b]})$. When parity even conditions are imposed on both $\sigma$ and $k_{ab}$, $\omega$ also has to be parity even and supertranslation charges identically vanish. This does not apply to translations $\omega = \zeta_{(\mu)}$, which are of odd parity, as these are still allowed and are associated with non-trivial charges. The supertranslations are therefore associated with non-vanishing charges only when the parity conditions \eqref{par5}-\eqref{par52} with $s_\s = s_k =1$ are not imposed as a boundary condition.

\vspace{3pt} \noindent {\bf Lorentz charges}\vspace{3pt}

The derivation of the Lorentz charges is much more involved. First, the property of asymptotic linearity
\bea
k_\xi [\delta g ; g ] = k_\xi [\delta g ; \bar g ], \label{al8}
\eea
does not hold for Lorentz transformations $\xi = -\xi_{(0)}$\footnote{We follow here the usual sign conventions of \cite{Iyer:1994ys}.}
. Moreover, the boundary fields vary as
\bea
\delta_\xi \s = -\mathcal L_{\xi_{(0)}} \s ,\qquad \delta_\xi k_{ab} = -\mathcal L_{\xi_{(0)}} k_{ab},
\eea
and the boundary counter-term charge does contribute. The computation goes as follows. The last term in \eqref{kk2} is explicitly zero for a rotation since $\xi_{(0)}=\xi_{(0),rot}$ is then tangent to the sphere and does not depend on the metric. This is the familiar result that the bulk charge associated with rotations is given by a Komar integral. However, in this case, the Komar integral contains a logarithmic divergence as well as a finite piece. For boosts $\xi_{(0)}=\xi_{(0),boost}$, the last term in \eqref{kk2} contributes and, therefore, the boost charge one-forms are not manifestly exact. It turns out to be manifestly exact only after we have substituted the Beig-Schmidt expansion in this expression. In intermediate expressions, a linearly divergent term in $\rho$ appears but is canceled between the Komar integral and the second term in \eqref{kk2}. The expression for the bulk charge eventually admits a logarithmic divergence and a finite piece as in the case of rotations.

The logarithmically divergent piece of the Lorentz charges takes the form
\bea
k_\xi [\delta g ; g ] = - \frac{\log\Lambda}{8\pi G} \int_S d^2S \sqrt{-h^{(0)}}\:  i_{ab} \: \xi^a_{(0)} n^b +O(\Lambda^0)\, .
\eea
It can be expressed using the linearization stability constraints \eqref{intFF} as
\bea
k_\xi [\delta g ; g ] =   - \frac{\log\Lambda}{4\pi G} \int_S d^2S \sqrt{-h^{(0)}} \Big(  T^{(\s)}_{ab}+ T^{(k)}_{ab} \Big)  \xi^a_{(0)} n^b +O(\Lambda^0),\label{infLo}
\eea
where $T^{(\s)}_{ab}$ and $T^{(k)}_{ab}$ are the stress-tensors of the actions \eqref{Ss}-\eqref{Sk}. Following our prescription, this divergence should be exactly canceled by the counter-term contributions to the total charge \eqref{totkk}. We have checked that the divergence indeed cancels after using the following relationship between symplectic structures and Noether charges,
\bea
\omega_{(\s)}( \delta \s,\mathcal L_{-\xi_{(0)}} \s) &=& \delta \Big( \sqrt{-h^{(0)}}T^{(\s)\:ab} \: \xi_{(0)\: b} (d^2x)_a \Big) - d^2S \sqrt{-h^{(0)}} n^a \DD^b (2 \xi_{(0)\: [a} \s_{b]}\delta \s)  ,\\
\omega_{(k)}(  \delta k, \mathcal L_{-\xi_{(0)}} k ) &=& \delta \Big( \sqrt{-h^{(0)}}T^{(k)\:ab} \: \xi_{(0)\:b} (d^2x)_a \Big) + d^2S \sqrt{-h^{(0)}} n^a \DD^b  P_{[ab]},
\eea
where $P_{ab}=P_{[ab]}$ is an anti-symmetric tensor.

Finally, the Lorentz charges are finite and have the following form
\bea
\mathcal Q_{-\xi_{(0)}}[g ;\bar  g ] &=& \frac{1}{8\pi G}\int_S d^2 S \sqrt{-h^{(0)}} \Big(
 - h^{(2)}_{ab} +\frac{1}{2} i_{ab} +\frac{1}{2} k_a^{c}k_{cb} \nn \\
&&+h_{ab}^{(0)} (8\s^2 +\s^c \s_c -\frac{1}{8}k_{ab}k^{ab}+k_{cd}\s^{cd} ) \Big) \xi_{(0)}^a n^b .\label{finalQrotb}
\eea
The charges are also conserved as a consequence of the momentum equation \eqref{eqh2c}. When parity conditions hold, the integral of all quadratic pieces vanish and the integral of $i_{ab}$ also vanishes as a consequence of the linearization stability constraints, see e.g., \cite{Mann:2008ay}. It then implies that asymptotic linearity \eqref{al8} holds and the counter-terms trivially integrate to zero. The charges thus agree with the Abbott-Deser formula
\bea
Q_{-\xi_{(0)}}[g ; \bar g] = \int_S  k_{-\xi_{(0)}} [ g-\bar g ; \bar g ] \, .
\eea
When $\sigma$ and $k_{ab}$ are not parity even, the charges \eqref{finalQrotb} contain quadratic terms in the fields and cannot be obtained from the linearized theory alone.

We showed in our previous work \cite{Compere:2011db} that the Lorentz charges can always be written in two equivalent ways using the conserved tensors $V_{ab}$ and $W_{ab}$ in the restricted phase space where $k_{ab} = i_{ab} = 0$. The tensor $V_{ab}$ can be related to the second order part of the boundary stress-tensor \cite{Mann:2008ay}. Let us remark here that, after using properties of integrals of tensor fields on the hyperboloid proven in Appendix B of \cite{Compere:2011db} and the equations of motion, we can rewrite the rotation and boost charges \eqref{finalQrotb} in the equivalent forms
\bea
\mathcal J_{(i)}&\equiv&\frac{1}{8\pi G}\int_S d^2S  \sqrt{-h^{(0)}} (V_{ab} +2 i_{ab}) \xi^a_{\rom{rot}(i)}n^b = - \frac{1}{8\pi G}\int_S d^2S \sqrt{-h^{(0)}}   W_{ab} \xi^a_{\rom{boost}(i)}n^b , \nn \\
\mathcal K_{(i)}&\equiv&\frac{1}{8\pi G}\int_S d^2S  \sqrt{-h^{(0)}}  (V_{ab}+2 i_{ab}) \xi^a_{\rom{boost}(i)}n^b = \frac{1}{8\pi G} \int_S d^2S  \sqrt{-h^{(0)}} W_{ab} \xi^a_{\rom{rot}(i)}n^b  \label{eq:K}.
\eea
where $V_{ab}$, $W_{ab}$ are defined in \eqref{defnlV}-\eqref{defnlW}. One can check \cite{Mann:2008ay} that for the Kerr black hole, one gets the standard result $\mathcal J_{(3)} = + m a$ for $\xi_{(0),rot}=\frac{\p}{\p \phi}$.

\vspace{3pt} \noindent {\bf Logarithmic translations}\vspace{3pt}

Logarithmic translations are allowed asymptotic transformations. They modify the first order fields as
\bea
\delta_\xi \s = H ,\qquad \delta_\xi k_{ab} = 0.
\eea
Since logarithmic translations do transform  $\sigma$, the boundary terms in the symplectic structure might play a role. Using integrations by parts one finds
\bea
H \s_b - H_b \s = -\frac{1}{2} E^{(1)}_{ab} H^a -  \DD^a (H_{[a}\s_{b]})
\eea
and therefore
\bea
\frac{ \log\Lambda}{4\pi G} \int_{S} \omega_{(\s)}( \delta \s,\delta_H \s )  =  \frac{\log \Lambda}{8 \pi G} \delta \int_S d^2 S \sqrt{-h^{(0)}} E^{(1)}_{ab} H^a n^b
\eea
where we have discarded the total divergence term on the sphere. The bulk covariant phase space charge associated with logarithmic translations is given by
\bea
\int_S k_\xi [\delta g ; g] = - \frac{\log \Lambda}{8 \pi G} \delta \int_S d^2 S \sqrt{-h^{(0)}} E^{(1)}_{ab}H^a n^b + \frac{1}{16 \pi G} \delta \int_S d^2 S  \sqrt{-h^{(0)}} k_{ab} n^a H^b \, .\label{infLog}
\eea
We find that the two divergent contributions are opposite of each other and exactly cancel. The remaining finite part is trivially integrable. Logarithmic translations are therefore associated with the non-trivial charges
\bea
\mathcal Q_{(H)} = \frac{1}{16 \pi G} \int_S d^2 S  \sqrt{-h^{(0)}} k_{ab} n^a H^b
\eea
which are conserved thanks to the property $\DD^a (k_{ab} H^b) = 0$. 

In the restricted phase space where $k_{ab} = 0$, logarithmic translations are associated with zero charges or equivalently they are degenerate directions of the symplectic structure. When parity conditions are imposed, logarithmic translations are not allowed transformations and the associated charges do not exist. The presence of non-vanishing conserved charges associated with logarithmic translations is therefore a particularity of the phase space without parity conditions and with $k_{ab} \neq 0$.

\subsection{Algebra of conserved charges}

In the last section, we obtained explicit expressions for conserved charges associated with translations, Lorentz transformations, logarithmic translations and supertranslations. We obtained that all asymptotic charges are non-trivial in general in our phase space. The set of infinitesimal diffeomorphisms form a Lie algebra defined from the commutator of generators. A natural question to ask is whether or not the algebra of translations,  logarithmic translations, Lorentz transformations and supertranslations is represented with the associated conserved charges.

General representation theorems are available \cite{Brown:1986ed,Barnich:2007bf} but one quickly realizes that they do not take into account boundary contributions to the symplectic structure. These contributions can be dealt with as follows. Every diffeomorphism in the bulk spacetime induces a specific transformation of the boundary fields through the Beig-Schmidt asymptotic expansion that identifies boundary fields from bulk fields. Therefore, the Lie algebra of infinitesimal diffeomorphisms defined from the commutator of generators also induces a Lie algebra of transformations of the boundary fields. The Poisson bracket between two charges is then defined as
\bea
\{ \mathcal Q_\xi [g ; \bar g], \mathcal Q_{\xi^\prime} [g ;\bar g] \} = - \delta_\xi \mathcal Q_{\xi^\prime} [g ;\bar g] ,\label{PB}
\eea
where the variation $\delta_\xi$ acts on the bulk fields as a Lie derivative and on the boundary fields as the transformation induced on the boundary fields from the Lie derivative of the bulk fields. It would be interesting to develop a general representation theorem which takes boundary contributions into account along the lines of \cite{Hollands:2005ya}. In this work, however, we simply evaluate the Poisson bracket using the explicit expressions for the charges derived in the previous section and taking into account the boundary field transformations.

Under an asymptotic translation $\xi = \omega(x) \p_\rho + o(\rho^{0})$ where $\DD_a \DD_b \omega +h_{ab}^{(0)}\omega = 0$, the boundary fields transform as
\bea
\delta_\omega \sigma &=& 0,\qquad \delta_\omega k_{ab} = 0,\\
\delta_\omega i_{ab} &=& 0, \qquad \delta_\omega V_{ab} = \DD_c (E^{(1)}_{ab}\omega^c ) + 2\eps_{cd (a}B^{(1)c}_{b)}\omega^d
\eea
where $E_{ab}^{(1)}$ and $B_{ab}^{(1)}$ are the first order electric and magnetic parts of the Weyl tensor while Lorentz transformations $\xi = -\xi_{(0)}$ act on the boundary fields as a Lie derivative
\bea
\delta_{-\xi_{(0)}} \sigma &=& \mathcal L_{-\xi_{(0)}}  \sigma ,\qquad \delta_{-\xi_{(0)}} k_{ab} = \mathcal L_{-\xi_{(0)}} k_{ab},\\
\delta_{-\xi_{(0)}} i_{ab} &=& \mathcal L_{-\xi_{(0)}}  i_{ab}, \qquad \delta_{-\xi_{(0)}} V_{ab} = \mathcal L_{-\xi_{(0)}} V_{ab} \,  .
\eea
Logarithmic translations act as
\bea
\delta_H \sigma &=& H,\qquad \delta k_{ab} = 0,\qquad 
\delta_H i_{ab}=  -\DD_c (E^{(1)}_{ab} H^c) -2 \epsilon^{mn}_{\:\:\:\:(a} B^{(1)}_{b)m} H_n \\
\delta_H V_{ab}&=& -\frac{1}{2} \DD_c (k_{ab} H^c)+ 2 \DD_c (E^{(1)}_{ab} H^c) + 8 \epsilon^{mn}_{\:\:\:\:(a} B^{(1)}_{b)m} H_n
\eea
and supertranslations act as
\bea
\delta_{\omega} \sigma &=&0  ,\qquad \delta_{\omega} k_{ab} = 2 \omega_{ab}+2 h_{ab}^{(0)}\omega, \qquad \delta_{\omega} i_{ab}  = 0 \\
\delta_{\omega} h^{(2)}_{ab} &=& k_{c(a}\omega_{b)}^{\; c} +k_{ab}\omega +\omega^c \left( \DD_c k_{ab} - \DD_{(a}k_{b)c}\right) \nn \\
&& + \Big( \s^c \omega_{c(ab)} - \s \omega_{ab}-2 \s \omega h_{ab}^{(0)} + \omega_{(a}\s_{b)} + \s_{c(a} \omega_{b)}^c + (\s \leftrightarrow \omega) \Big) .
\eea

After an explicit evaluation, we find that all asymptotic transformations are well-represented: the Poisson bracket is anti-symmetric and is isomorphic to the semi-direct product of the Lorentz algebra with the (super)-translation and logarithmic translation algebra. In particular, the Poisson bracket between Lorentz charges and (super)-translation charges is given by
\bea
\{ \mathcal Q_{-\xi_{(0)}} , \mathcal Q_{(\omega)} \} = - \{   \mathcal Q_{(\omega)} , \mathcal Q_{-\xi_{(0)}}\} = \mathcal Q_{(\omega^\prime)} ,\qquad \omega^\prime = \mathcal L_{-\xi_{(0)}}\omega\, .
\eea
and the Poisson bracket between Lorentz charges and log-translation charges is
\bea
\{ \mathcal Q_{-\xi_{(0)}} , \mathcal Q_{(H)} \} = - \{   \mathcal Q_{(H)} , \mathcal Q_{-\xi_{(0)}}\} = \mathcal Q_{(H^\prime)} ,\qquad H^\prime = \mathcal L_{-\xi_{(0)}}H\, .
\eea
Logarithmic translations and supertranslations obey the algebra
\beqs 
 \{ \Q_{(\omega)} ,  \Q_{(H)} \} = - \{ \Q_{(H)} , \Q_{(\omega)} \} = \frac{1}{4\pi G} \int d^2 S \: \sqrt{-h^{(0)}} \: n_a \Big( H^a \omega - H \omega^a \Big)
\eeqs
where the right-hand side depends on the generators but does not depend on the fields. In the harmonic decomposition of $\omega$ on the sphere, the  Poisson bracket is zero for all harmonics $ l >1$ and is a Kronecker delta for the four lowest harmonics $l \leq 1$. The algebra of asymptotic conserved charges is isomorphic to the algebra of asymptotic symmetries. No non-trivial central extension of the algebra is present.

In order to derive these results, we used Appendix \ref{appSD} and the Appendix B of \cite{Compere:2011db} to simplify intermediate expressions and we discarded boundary terms. We have also used the property described in \cite{Ashtekar:1978zz,BS} (see also Appendix \ref{app:Nut}) that regularity of $k_{ab}$ implies that the four conserved NUT charges \cite{Nutdef}
\bea
\mathcal P_{(\mu)} = \frac{1}{8\pi G}\int_S d^2 S\, B_{ab}^{(1)} n^a \DD^b \zeta_{(\mu)},\qquad \mu=0,1,2,3,\label{noNut}
\eea
are zero. The presence of NUT charges would require modifications of the Poisson bracket, see e.g. \cite{Argurio:2008zt} where a symmetric topological extension of the algebra of bosonic supercharges is considered. The fact that we find an anti-symmetric Poisson bracket gives us confidence in our intermediate expressions. 

We obtained that all transformations: translations, logarithmic translations, Lorentz transformations and supertranslations are well-represented despite the (log/super)-translation anomaly. The fact that the Lorentz group is well-represented is not surprizing given that the cut-off needed to regularize the action, see \eqref{action}, is invariant under asymptotic Lorentz transformations.  Now, it is also important to take into account the shifts in the action when one changes the cut-off used to regulate the action. These shifts can be analyzed as follows. 

Under a change of cut-off $\Lambda$, the action will be shifted by a finite piece $S_{(0)}$ proportional to the anomaly action $ S^{(\s)} + S^{(k)}$ given in \eqref{Ss}-\eqref{Sk}. The conserved charges associated with the asymptotic Killing vector $\xi$ will then be shifted by the boundary Noether charges of the action $S^{(\s)} + S^{(k)}$ associated with the symmetry $\delta_\xi$. Using standard manipulations $\delta_\xi L = d K_\xi$, $\delta L = \frac{\delta L}{\delta \phi}\delta \phi + d \Theta [\delta \phi ]$, the boundary Noether charges are defined as $J_\xi = K_\xi - \Theta[\delta_\xi \phi]$. One then quickly sees that translations and supertranslations are associated with vanishing Noether charges $\int_S d^2 S J_\xi = 0$ while Noether charges associated with logarithmic translations are proportional to the four-momentum $\mathcal Q_{(\mu)}$ and  Noether charges associated with Lorentz transformations are given by the integral of $J_{\xi^{(0)}} = 2(T^{(\s)ab} + T^{(k) ab} )\xi^{(0)}_b (d^2 x)_a$ where $T^{(\s)ab}$ and $T^{(k)ab}$ are the stress-tensors of the actions \eqref{Ss}-\eqref{Sk}. 

Therefore, under a change of regulator, the translations and supertranslation charges are invariant. Logarithmic translation charges get shifted with the four-momenta and the Lorentz charges get shifted as
\bea
\Delta \mathcal Q_{-\xi_{(0)}}[g ;\bar  g ] \sim   \int_S d^2S  \sqrt{-h^{(0)}} ( T^{(\s)}_{ab} + T^{(k)}_{ab} ) \xi^a_{(0)} n^b\, .
\eea
These shifts can be obtained similarly by varying the regulator directly into the expression for the charges \eqref{infLo}-\eqref{infLog} before the subtraction of divergences between the bulk and the boundary. Four-momenta and supertranslations are finite without needing a regulator. They are therefore manifestly unchanged by the regulator.

\section{Hamiltonian dynamics}
\label{sec:H}

We have seen that parity conditions on the hyperboloid are not required in order to define a consistent phase space in the hyperbolic representation of spatial infinity. Moreover, we have seen that when these conditions are relaxed, charges associated with Lorentz rotations and boosts are non-linear functionals of the first order fields and logarithmic translations and supertranslations are associated with non-trivial charges. Both these characteristics are not shared with the standard treatment of Hamiltonian charges at spatial infinity \cite{Arnowitt:1962aa,Regge:1974zd,Beig:1987aa}. There, parity conditions on the sphere are imposed in order that the rotation and Lorentz boost charges are finite. Also, charges are linear functionals of the boundary fields, logarithmic translations are not allowed transformations and supertranslations are associated with trivial Hamiltonian generators. The purpose of this section is to resolve this tension by proposing how the results of \cite{Regge:1974zd,Beig:1987aa} can be accommodated to enlarge the phase space to fields which do not obey parity conditions.

The canonical two-form on the canonical phase space used in the treatments of \cite{Arnowitt:1962aa,Regge:1974zd,Beig:1987aa} is the bulk canonical two-form
\bea
\Omega(\delta_1 \g,\delta_1 \pi,\delta_2 \g,\delta_2 \pi ) = \frac{1}{16 \pi G} \int_\Sigma d^3 x\Big(  \delta_1 \pi^{mn} \delta_2 \g_{mn} - \delta_2 \pi^{mn} \delta_1 \g_{mn} \Big)\,
\eea
defined from the bulk canonical fields $ \g_{mn}$ and $\pi^{mn}$ at the initial time surface $\Sigma$ at $t=0$. In the case of asymptotically flat spacetimes without parity conditions the bulk canonical two-form suffers from a logarithmic radial divergence. Using the boundary conditions \eqref{geng1}-\eqref{geng2} one can express the canonical two-form as
\bea
\Omega(\delta_1 \g,\delta_1 \pi,\delta_2 \g,\delta_2 \pi ) = (finite)  + \frac{\log \Lambda}{16 \pi G} \int_{S} d^2 S \Big( \delta_1 \pi^{(1)\: mn} \delta_2 \g^{(1)}_{mn} - \delta_2 \pi^{(1)\: mn} \delta_1 \g^{(1)}_{mn} \Big)\label{can1}
\eea
where $\Lambda$ is a large radial cut-off and $S$ is the sphere at $r = \Lambda$. Now,  in complete parallel to the Lagrangian treatment, we propose to modify the dynamics by adding a boundary term to the canonical form. We proceed by first writing the boundary actions \eqref{Ss}- \eqref{Sk} at $t=0$ in the 2+1 decomposition (the boundary metric becomes the real time line times the unit sphere). We then switch to the Hamiltonian formulation of the boundary action and propose to supplement the bulk canonical fields with the canonical fields of the boundary Hamiltonian. We then introduce counter-terms to the canonical form in order to minimally cancel the divergences in \eqref{can1},  in complete parallel to the Lagrangian prescription \eqref{Omegact}. The regulation breaks translation, supertranslation and logarithmic translation invariance. We interpret this breakdown as a consequence of the translation anomaly in the action, which is manifest only when fields have both parities.

Let us now discuss briefly the form of the Hamiltonian generators associated with asymptotic Poincar\'e transformations and supertranslations. The Hamiltonian generators contain two parts: the part coming from the bulk canonical form and the counter-term contribution that cancels the logarithmic divergences. The surface charge derived from the bulk canonical form associated with a gauge parameter $\eps^A = (\eps^\perp,\eps^m)$ is given by \cite{Regge:1974zd}
\bea
k_\eps^{[0m]}[\delta \g,\delta \pi ; \g,\pi] = G^{mnop}(\eps^\perp \delta \g_{op|n} - \eps^\perp_{|n} \delta \g_{op}) + 2 \eps^o \delta (\g_{on}\pi^{mn}) - \eps^m \delta\g_{no}\pi^{no} \label{RT}
\eea
where
\bea
G^{mnop} = \frac{1}{2}\sqrt{\g} \left( \g^{mo}\g^{np} +\g^{mp}\g^{no} - 2\g^{mn}\g^{op} \right)
\eea
is the inverse De Witt supermetric. Now, one can readily obtain that this expression admits at most a logarithmic divergence when one uses our boundary conditions. The explicit asymptotic expansions provided in Appendix \ref{app:BC} are useful in deriving the explicit forms of the charges. The logarithmic divergence is then exactly canceled by the boundary counter-term. The resulting final expressions for the charges in Hamiltonian formalism can then be obtained by a straightforward explicit evaluation. We will not provide them here. We only note that the four-momenta are given by the usual ADM formulae, while the charges associated with rotations and boosts contain non-linear contributions in the canonical fields.

\section{Summary and Discussion}
\label{sec:D}

We showed that the four-dimensional Einstein-Hilbert action admits a logarithmic divergence at spatial infinity when one regulates the action using hyperbolic temporal slices or, in other words, when one considers the action principle between initial and final time slices that are relatively boosted with respect to each other at spatial infinity. We described two alternative mechanisms to cancel the divergence: either one imposes parity conditions on the first order boundary fields or one adds to the action an anomalous counterterm that breaks asymptotic diffeomorphism invariance including asymptotic translation invariance. We will summarize what each choice of boundary conditions implies.

If one insists in requiring asymptotic Poincar\'e invariance, one is led to impose parity conditions. In that case, we built a bigger covariant phase space than the one previously considered \cite{Ashtekar:1978zz,ABR} while preserving all essential properties of asymptotic flatness such as the Poincar\'e group as asymptotic symmetry group and the inexistence of other conserved charges at spatial infinity than the Poincar\'e charges. The enhancement of the covariant phase space that we have constructed with respect to the one of \cite{Ashtekar:1978zz,ABR} originates from the admissibility of fluctuations of the first order part of the Weyl tensor (while still imposing regularity of all fields and therefore no NUT-momenta) and from the presence of temporal and radial supertranslations that are pure gauge. We will further comment on this generalized phase space below.

If one insists in building the largest possible phase space, one is led to relax the parity conditions. This set of asymptotically flat spacetimes has new qualitative features that we unraveled. For such relaxed boundary conditions, the existence of a variational principle requires to add an anomalous counterterm to the Einstein-Hilbert action. This counterterm arises as follows. Logarithmic divergent integrals at the spatial boundary of temporal hyperbolic initial and final slices in the action are regulated by introducing  a hyperbolic radial slicing and by cancelling divergences at a finite radial hyperbolic cutoff. Thanks to the choice of hyperbolic cutoff, the regulation procedure preserves asymptotic Lorentz invariance. If this cutoff is pushed to infinity, the resulting regulated action will not be invariant under asymptotic radial diffeomorphisms since the action would be shifted by a finite piece. This finite action is precisely the anomaly and is proportional to the action for the first order boundary fields that one can infer from the bulk Einstein's equations. Therefore, the regulated action depends on the specific choice of hyperbolic radial foliation close to spatial infinity. A covariant phase space is defined only after one such slicing has been chosen and fixed. The anomalous counterterm is given by the anomaly multiplied by the logarithm of the radial hyperbolic cutoff. The action for asymptotically flat spacetimes without parity conditions has a radial hyperbolic slicing anomaly in the same sense that odd-dimensional anti-de Sitter spacetimes have a Weyl anomaly \cite{Henningson:1998gx}\footnote{The presence of anomalies is uncorrelated with the presence of central charges in the asymptotic symmetry group. Here, as in anti-de Sitter spacetimes in odd spacetime dimensions $d \geq 5$, the asymptotic symmetry group is finite-dimensional and not centrally-extended while anomalies are present.}. The interpretation of this anomaly and the role that it may play in the quantization of asymptotically flat gravity largely remains to be understood. 

As a first step in the understanding of the structure of asymptotically flat spacetimes without parity conditions, we derived the asymptotic charges and the asymptotic symmetry group using the details of the Beig-Schmidt expansion. The hyperbolic radial slicing is not invariant under the action of (super/log)-translations. As expected from general considerations, the anomaly is however invariant under (super/log)-translations and Lorentz transformations. The relaxed boundary conditions lead to several additional non-vanishing conserved charges at spatial infinity such as logarithmic translation charges, supertranslation charges and boundary Noether charges. The logarithmic translation and supertranslation charges are canonically associated with bulk (super/log)-translations that are however not allowed to act on the phase space. The boundary Noether charges are associated with the boundary Lorentz symmetries of the anomaly. When parity conditions do not hold, the Lorentz charges are non-linear functionals of the asymptotic fields and therefore differ from the standard ADM and AD formulas \cite{Arnowitt:1962aa,Abbott:1981ff}. The standard ADM and AD formulas are restored when parity conditions hold. The non-linearities have been treated in detail in this work. Such asymptotic non-linearities appeared already for asymptotically anti-de Sitter spacetimes. In Einstein gravity, the charges are linear functionals of field perturbations around anti-de Sitter \cite{Abbott:1981ff,AM1984ADS,Henneaux:1985tv}. However, non-linearities may appear when matter fields are present, see e.g., \cite{Bianchi:2001de,Bianchi:2001kw,Henneaux:2006hk}.

It is interesting to remark that the presence of non-vanishing charges associated with supertranslations in addition to Poincar\'e transformations is also a feature of null infinity where supertranslations along the null direction are also associated with non-trivial charges \cite{BMS,DrayStreubel,Wald:1999wa}\footnote{The proposal of considering the so-called superrotations \cite{Barnich:2010eb,Barnich:2011ct} in addition to supertranslations at spatial infinity would also require a careful regularization of the super Lorentz charges, see \cite{Barnich:2011mi}, which would go beyond the considerations of this work.}. For regular asymptotic fields, one expects that supertranslations charges should be conserved at infinite past times of future null infinity or at infinite late times of past null infinity where the news tensor vanishes. Indeed, at such late or early times the expression of \cite{BMS,DrayStreubel,Wald:1999wa} becomes conserved and proportional to the first order electric part of the Weyl tensor and matches qualitatively with our expression \eqref{Qsupertransl}. The shifts of Lorentz charges due to a change of radial hyperbolic cutoff are also reminiscient of the ambiguities in defining Lorentz charges at null infinity due to the action of supertranslations. It would be interesting to make that qualitative agreement more precise by comparing the precise definitions of supertranslations.

We derived the asymptotic symmetry group using a generalization of the Poisson bracket taking into account boundary contributions to the charges. We obtained that the Poisson bracket is well-defined for all asymptotic symmetries: translations, logarithmic translations, supertranslations and Lorentz transformations. We derived the complete asymptotic symmetry algebra of conserved charges and we obtained that it is not centrally extended. The situation here can be contrasted to bulk infinitesimal diffeomorphisms which induce Killing symmetries and conformal Killing symmetries of asymptotically AdS spacetimes in odd dimensions as analyzed in \cite{de Haro:2000xn,Skenderis:2000in,Hollands:2005ya,Papadimitriou:2005ii}. First, the dependence of the Lorentz charges, associated with Killing vectors, upon the choice of regulator is analogous to the shift of the stress-tensor by Weyl anomalous terms \cite{de Haro:2000xn,Skenderis:2000in}. In our case, logarithmic translations are also present and they are also shifted under a change of regulator. In anti-de Sitter, infinitesimal diffeomorphisms associated with boundary conformal Killing vectors are well-represented by the Poisson bracket even though they may act non-trivially on the action \cite{Hollands:2005ya}. Indeed, the action only varies by a c-number which depends on the boundary conditions while the dynamical phase space is preserved. The non-conservation of the associated charges is related to this c-number. In asymptotically flat spacetimes, translations are also boundary conformal Killing vectors. Four-momenta as well as supertranslations are always exactly conserved and they do not vary under a change of regulator.

No exact solution of vacuum Einstein's equations is known to us which breaks parity conditions. Such a solution would possess twelve boundary Noether charges in addition to Poincar\'e, logarithmic translation and supertranslation charges. The boundary Noether charges are the Noether charges of the actions for the first order fields associated with the boundary Killing symmetries or equivalently with the asymptotic Lorentz Killing vectors. A subclass of those solutions exists as an analytic series expansion at spatial infinity. Indeed, one can consistently set the logarithmic terms in the expansions \eqref{metric1} and \eqref{geng1sub}-\eqref{geng2sub} to zero and still obey Einstein's equations by fixing six linear combinations of the boundary Noether charges to zero, see Section \ref{sec:LS} for details. Then, the original Beig-Schmidt expansion \cite{BS} which uses only polynomials in $\rho$ is a consistent analytic asymptotic solution of Einstein's equations at all asymptotic orders which has six boundary Noether charges. We leave the existence, or not, of a regular solution in the bulk with such charges as an open question.

As a side note, we could also generalize the construction of asymptotically flat spacetimes to include NUT charges. It would require to allow the field $k_{ab}$ to contain Dirac-Misner string singularities. We present in Appendix \ref{app:Nut} a preliminary lemma useful in that context. As a consequence of the lemma, the NUT four-momenta can be defined as the surface integral of the first order magnetic part of the Weyl tensor as \eqref{noNut}, in complete analogy with the four-momenta \eqref{ch:tr}, defined as an integral of the first order electric part of the Weyl tensor. The asymptotic symmetry group would be modified, see also  \cite{Argurio:2008zt,Argurio:2008nb,Argurio:2009xr}. Due to singularities in $k_{ab}$, the definition of Lorentz charges would require a careful treatment of integrals on the sphere, especially the boundary terms on the sphere that we ignored in this work, which would make the analysis technically much more involved.

We started this work by emphasizing the following puzzle: how to reconcile the two facts that parity conditions do not seem to be required for defining a variational problem for asymptotically flat spacetimes while parity conditions are required for defining the covariant phase space symplectic structure and canonical two-form. We found that the puzzle can be resolved in the framework of regulating asymptotically flat spacetimes with hyperbolic cutoffs 
by considering the boundary terms in the variational principle on hyperbolic temporal cutoffs. When these boundary terms are considered, we showed that a variational principle which allows for asymptotic Poincar\'e diffeomorphisms exists only when parity conditions are imposed. This argument was argued to be independent on the particular counterterms chosen at spatial infinity, e.g., in the choice of the Mann-Marolf prescription. 

We stressed in the introduction that supertranslations are not on the same footing between the canonical formalism, where some supertranslations are allowed to act on the fields but are pure gauge, and the standard treatment of the covariant phase space where supertranslations are completely fixed. We partially resolved that tension by constructing a covariant phase space with parity conditions on the first order fields that admits some supertranslations that are pure gauge. This covariant phase space enlarges the one defined in \cite{Ashtekar:1978zz,ABR} by allowing to vary the dynamical field $k_{ab}$ (restricted to be traceless, divergence-free and regular). This phase space therefore admits non-trivial first order part of the magnetic Weyl tensor $B_{ab}^{(1)} \neq 0$ while still having zero NUT charge. 
We showed that a sufficient condition to have a well-defined variational principle on this phase space is to fix the trace and divergence of $k_{ab}$ while using the Mann-Marolf variational principle. Interestingly, supertranslations that leave the trace of $k_{ab}$ invariant are precisely the supertranslations that are conserved on the phase space. Also, these supertranslations are precisely the covariant analogue of temporal and radial supertranslations in the canonical formalim \cite{Regge:1974zd}. We only consider these supertranslations in this work. Logarithmic translations and parity-odd supertranslations do not act on this phase space because of the parity conditions. Parity-even supertranslations act on the phase space but are pure gauge. This covariant phase space has all properties that one expects such as the Poincar\'e group as asymptotic symmetry group. The anomaly, even though non-vanishing off-shell, does not contribute to the classical dynamics as a consequence of the parity conditions. Also, Poincar\'e charges are linear functionals of asymptotic fields and reduce to the standard expressions for conserved charges. An auxiliary question remains on the role of the so-called angular supertranslations (see Section \ref{sec:BC} for details), which can be summarized as follows. In the Hamiltonian boundary conditions stated in \cite{Regge:1974zd}, angular supertranslations are allowed transformations and are pure gauge while these diffeomorphisms are not allowed in the covariant phase space where one fixes the mixed components $g_{\rho a}$ in the Beig-Schmidt expansion. The introduction of mixed components $g_{\rho a}$ in the Beig-Schmidt expansion would therefore be necessary in order to fully obtain the Lagrangian analogue of the Hamiltonian boundary conditions stated in \cite{Regge:1974zd}. We leave such a generalization for future work. %Also, since temporal and radial supertranslations are associated with non-trivial charges when parity conditions do not hold, it would be interesting to check whether or not angular supertranslations are associated with finite charges when parity conditions do not hold. 

The second puzzle that we raised in the introduction was related to the status of log-translations and parity-odd supertranslations: should they be considered as unphysical diffeomorphisms (e.g. are they gauge transformations in an enlarged phase space) or physical diffeomorphisms (are they associated with conserved charges)? We attempted to resolve this puzzle by constructing a phase space where these transformations are allowed diffeomorphisms. In fine, we only constructed a collection of phase spaces without parity conditions where a radial hyperbolic foliation is required to uniquely define each phase space. Therefore, even for such generalized phase spaces, log-translations and parity-odd supertranslations are not allowed to act on the phase space because of the anomaly shift of the action under such a transformation. However, the diffeomorphisms are associated with non-vanishing conserved charges in general. The puzzle therefore remains. Lacking a clear interpretation of the covariant phase spaces without parity conditions, we leave to further work the elucidation of the role of these asymptotic diffeomorphisms. More generally, it is not clear what the anomaly tells us about the quantization of Einstein gravity. Nevertheless, since the anomaly can be derived from first principles from the Einstein-Hilbert action under reasonable assumptions, we think that it may find its place in (and maybe contribute to formulate) a future theory of quantum gravity in asymptotically flat spacetimes.

\subsection*{Acknowledgements}
We gratefully acknowledge A. Ashtekar, G. Barnich, D. Marolf, K. Skenderis, C. Troessaert, A. Virmani and B. Wouters for enlightning discussions. We thank especially A. Virmani for a productive collaboration at early stages of this project and for a careful reading of the manuscript, A. Ashtekar, D. Marolf and our referee for their sharp suggestions and B. Wouters for pointing to one omitted term in the variation of the action in an earlier version of this manuscript. GC acknowledges support from a `Nederlandse Organisatie voor Wetenschappelijk Onderzoek' (NWO) Vici grant. FD is supported by IISN Belgium (conventions 4.4511.06 and 4.4514.08), and by the Belgian Federal Science Policy Office through the Interuniversity Attraction Pole P6/11.

\appendix

\section{Comparison of 3+1 and covariant boundary conditions}
\label{app:BC}

The hyperbolic and cylindrical representation of spatial infinity are valid in the limits $\rho \rightarrow \infty$ and $r \rightarrow \infty$, respectively. The key change of coordinates is the one mapping flat spacetime from the hyperbolic to the cylindrical representation of spatial infinity
\bea
\rho = r \sqrt{1-\frac{t^2}{r^2}},\qquad \tau = \text{arctanh}{(\frac{t}{r})}\, .
\eea
The hyperbolic and cylindrical representations coincide asymptotically in the limit where ADM time is kept finite, $t / r \rightarrow 0$ which is equivalent to $\tau \rightarrow 0$. In that case, $\rho \sim r$ asymptotically.

In order to obtain the form of the metric in $r,t$ coordinates, we expand the right-hand side of $\rho,\tau$ in powers of $t/r$ and we expand the Beig-Schmidt fields in Taylor series around $\tau = 0$,
\bea
\s(\tau,\th,\phi) &=& \s(\th,\phi) + \frac{t}{r} \s^\pi(\th,\phi) + \frac{t^2}{2 r^2} \gamma(\th,\phi) + O(r^{-3}) ,\\
k_{ab}(\tau,\th,\phi) &=& k_{ab}(\th,\phi) + \frac{t}{r} k^\pi_{ab}(\th,\phi) + \frac{t^2}{2 r^2} \gamma_{ab}(\th,\phi) + O(r^{-3}) ,\\
i_{ab} (\tau,\th,\phi) &=& i_{ab}(\th,\phi) + \frac{t}{r}  i^\pi_{ab}(\th,\phi) + O(r^{-2}),\\
h^{(2)}_{ab} (\tau,\th,\phi) &=& h^{(2)}_{ab}(\th,\phi) + \frac{t}{r}  h^{\pi,(2)}_{ab}(\th,\phi) + O(r^{-2}),
\eea
where we define $\s^\pi(\th,\phi)= \p_\tau  \s(0,\th,\phi)$, $k_{ab}^\pi(\th,\phi)= \p_\tau  k_{ab}(0,\th,\phi)$, $ \gamma(\th,\phi) = \p_\tau \p_\tau \s(0,\th,\phi)$, $ \gamma_{ab}(\th,\phi) = \p_\tau \p_\tau k_{ab}(0,\th,\phi)$, $i^\pi_{ab} =\p_\tau i_{ab}(0,\th,\phi)$, $h^{\pi,(2)}_{ab}(\th,\phi) = \p_\tau h^{(2)}_{ab}(0,\th,\phi)$. We will keep the same notation for canonical fields in Hamiltonian formalism as fields in Lagrangian formalism
\bea
\s(0,\th,\phi) = \s(\th,\phi), \qquad k_{ab} (0,\th,\phi) = k_{ab} (\th,\phi) , \nn \\
i_{ab} (0,\th,\phi) = i_{ab} (\th,\phi),\qquad h^{(2)}_{ab} (0,\th,\phi) = h^{(2)}_{ab} (\th,\phi) .
\eea
%We hope not to confuse the reader.
The tensors decompose into scalars, vectors and two-dimensional tensors under decomposition into temporal and spatial components. The meaning of the notation should be clear in either Hamiltonian or Lagrangian context. The fields $\gamma(\th,\phi)$ and $\gamma_{ab}(\th,\phi)$ are
 determined from the equations of motion of $\sigma$ and $k_{ab}$. After a straightforward computation, we obtain
\bea
\g_{rr} &=& 1 + \frac{2\s}{r} + \frac{\s^2 +2 t \s^\pi }{r^2} +o(r^{-2}  ) ,\nn \\
\g_{r \zeta} &=& - \frac{t}{r}k_{\tau \zeta} - t \frac{\log r (i_{\tau\zeta}) + h^{(2)}_{\tau \zeta} + t k^\pi_{\tau \zeta} }{r^2} + o(r^{-2}) \label{g2BC} ,\\
\g_{\zeta \iota} &=& r^2 g_{\zeta \iota} +(k_{\iota \zeta} - 2 \s  g_{\zeta \iota} ) r+ \log r (i_{\zeta\iota} )+ (h^{(2)}_{\zeta \iota}+ t k^\pi_{\iota \zeta} -2 t \s^\pi g_{\zeta \iota}) +o(r^0), \nn
\eea
for the canonical fields and
\bea
(det g_{\zeta \iota})^{-1/2}\pi^{rr} &=& -2 \s^\pi + \frac{1}{2}k^\pi_{\iota \zeta} g^{\iota \zeta}_{(S^2)}- D_{(S^2)}^\iota k_{\tau \iota} + \frac{\log r}{r} \Big( \frac{1}{2} g^{\zeta \iota} i^\pi_{\zeta\iota} -   D_{(S^2)}^\zeta i_{\tau \zeta}\Big) \nn \\
&&+ \frac{1}{r} \Big( \frac{1}{2} g^{\zeta \iota} h^{\pi,(2)}_{\zeta \iota} - D_{(S^2)}^\zeta h^{(2)}_{\tau \zeta}  -  2 t \gamma + 6 t \s -\frac{1}{2}t k_{\iota\zeta} g_{(S^2)}^{\iota \zeta} +\frac{1}{2} t \gamma_{\iota \zeta} g^{\iota\zeta}_{(S^2)} \nn\\
&& + 2 t k_{\tau\tau} - k_{\tau\tau}\s^\pi - t D_{(S^2)}^\iota k^\pi_\iota + (k-k \; \text{terms})\Big)  + o(r^{-1}),\nn \\
(det g_{\zeta \iota})^{-1/2}\pi^{r \iota} &=& -\frac{1}{2 r}k_\tau^\iota - \frac{\log r}{r^2} (i_{\tau}^{\iota} ) +\frac{1}{r^2} \Big(-h^{(2)\iota }_{\tau} + \frac{1}{2}i_{\tau}^{\iota} -2 t \p^\iota \s \label{g2BC2} \\
&& -\frac{t}{2}k^\pi_{\tau \zeta} g^{\zeta\iota}_{(S^2)}-\s k_{\tau \zeta} g^{\zeta\iota}_{(S^2)} - \frac{t}{2} D_{(S^2)}^\iota k_{\tau\tau} + (k-k \; \text{terms})  \Big) + o(r^{-2}) ,\nn \\
(det g_{\zeta \iota})^{-1/2}\pi^{\iota \zeta} &=& \frac{1}{r^2} \Big( - \frac{1}{2} k^{\pi, \iota \zeta} + D_{(S^2)}^{(\iota} k^{\zeta)}_\tau +g^{\iota \zeta}_{(S^2)}(\frac{1}{2}k^{\pi,\xi}_\xi - D_{(S^2)}^\xi k_{\tau \xi})  \Big) \nn\\
&&+ \frac{\log r}{r^3} \Big( - \frac{1}{2} i^{\pi, \iota \zeta} + D_{(S^2)}^{(\iota} i^{\zeta)}_\tau +g^{\iota \zeta}_{(S^2)}(\frac{1}{2}i^{\pi,\xi}_\xi - D_{(S^2)}^\xi i_{\tau \xi})  \Big) + O(r^{-3}) ,\nn
% \\ && + \frac{1}{r^3} (....) + o(r^{-3}) \nn
\eea
for the conjugate fields. Here, we denote by $(k-k \;\text{terms})$ terms quadratic in $k_{ab}$ which do contribute to the finite part of the conserved Lorentz charges but that we omit here for simplicity.

Let us finally discuss how the notions of parity are related between Beig-Schmidt fields and canonical fields. A field on the hyperboloid is parity-time reversal even if it is invariant under the combined transformation of inverting the hyperboloid time $\tau \rightarrow -\tau$ and doing a parity transformation $(\theta,\phi)\rightarrow (\pi-\theta,\phi+\pi)$. Fields in canonical formalism are parity-time reversal even if their components in Cartesian coordinates do not transform under three-dimensional parity and if the components of their conjugate momentum in Cartesian coordinates transform with an overall sign under parity. From the dictionary of the Beig-Schmidt asymptotic fields in 3+1 decomposition,  we see after switching from spherical to Cartesian coordinates that the even parity-time reversal conditions on $\sigma$ and $k_{ab}$ lead to parity-time reversal even first order canonical fields on the initial time slice $t=0$.

\section{Classification of symmetric and divergence-free tensors}
\label{appSD}
\label{sec:SD}

As discussed in the main text, conserved quantities associated with Poincar\'e generators can be built out of symmetric divergence-free and traceless (SDT) tensors, or more generally out of symmetric and divergence-free (SD) tensors. It is useful in order to establish unicity of the definition of conserved charges and in order to understand the structure of the  linearization stability constraints at second order to provide the classification of all possible SD tensors that one can built up from quadratic terms in $\sigma$ and $k_{ab}$ and their derivatives. In fact, for both purposes, it is sufficient to classify SD tensors whose curl are non-zero since the integral on the sphere of a SD tensor with zero curl contracted with a Killing or conformal Killing vector is identically zero (see Appendix B of \cite{Compere:2011db}).

All SD tensors built out of quadratic terms in the first order fields can be formed from symmetric tensors $M_{ab}$ obeying $\DD^b M_{ab} = \DD_a M$ and which we call tensor potentials. A complete set of SD tensors consists of SD tensors given by $\kappa_{ab}= M_{ab}-M h^{(0)}_{ab}$ and of SDT tensors obtained by acting with successive curls on $M_{ab}$ or equivalently by acting with successive symmetrized curls on $\kappa_{ab}$. Indeed, an SDT tensor can be constructed from $T_{ab} = \eps_a^{\;\, cd}\DD_c M_{d b }= \eps_{cd(a}\DD^c \kappa^{\:\:\: d}_{b)}$. As we have just emphasized, the curl of a tensor potential might be trivially zero. Therefore the SD and SDT tensors whose curl are non-zero will turn out to be classified using the equivalence of classes of tensor potentials where two tensor potentials are equivalent if their difference has a trivial curl. We will denote one representative of such equivalence class of non-trivial tensor potentials a RNT tensor potential.

We explain in the next subsection the procedure we have followed in order to obtain and prove that we have obtained all RNT tensor potentials needed to construct any SD or SDT tensor whose curl is non-zero.

\subsection{Algorithm for classification}

The first order fields $\sigma$ and $k_{ab}$ obey decoupled linear equations. We thus consider separately the quadratic combinations $(\sigma,\sigma)$, $(k,k)$ and $(\sigma,k)$. For each case, we use the following procedure:

\begin{enumerate}

\item We start by listing a basis of rank two symmetric tensors with $m$ derivatives built out of quadratic terms which are independent on-shell. It exists a number of derivatives $m^\star$ such that for all $m \geq m^\star$ the number of terms in that basis is maximal,  i.e. no new tensor structure appears at higher order. At lower values $m < m^\star$ not all possible tensor structures can appear due to a lack of derivatives. We find $m^\star = 3$ for $(\s,\s)$, $m^\star = 3$ for $(k,k)$ and $m^\star = 4$ for $(k,\s)$ tensors. Due to the presence or not of the epsilon tensor depending if $m$ is even or odd, the general form of a rank two symmmetric tensor built out of linear combinations of the basis takes a different form. We denote this tensor as $Q^{(2n)}_{ab}$ or $Q^{(2n+1)}_{ab}$ and we provide its general form.

\item We continue by deriving a bound on the possible SDT tensors that one can build at a fixed number $m \geq m^\star$ of derivatives. We simply compute the number $H$ of linearly independent tensors $Q^{(m)}_{ab}$ which obey both $\DD^b Q^{(m)}_{ab}=0$ and $Q^{(m)a}_a=0$ where equalities here are valid up to terms with lower derivatives. At this stage, the number $H$ is only a bound on the number of SDT tensors at order $m$ because none of them has been constructed fully yet. We obtain that $H=1$ in the  $(\sigma, \sigma)$ case, $H=3$ in the $(k,k)$ case and $H=2$  in the $(\sigma, k)$ case for both $m$ even or odd.

\item We then derive the explicit form of all RNT potentials, SD tensors and SDT tensors at each low value $m \leq m^\star$ of derivatives by enumeration. We write a basis of rank two symmetric tensors built out of quadratic terms which are independent on-shell with at most $m$ derivatives for each $m \leq m^\star$ and impose the RNT or SDT conditions. The SD tensors $\kappa_{ab}$, whose curls are non-zero, are obtained from the RNT potentials by the correspondence $\kappa_{ab} = M_{ab} - h_{ab}^{(0)} M^c_c$.

\item We finally observe that there are exactly $H$ SDT tensors which have at most $m^\star$ derivatives and at least one term with $m^\star$ derivatives.  This provides a proof that each candidate SDT tensor exists at order $m^\star$. We then note that the SDT tensors obtained by acting with the curl operator on these tensors form a basis for SDT tensors at order $m^\star + 1$ and by successive iterations at each order $m \geq m^\star$. Since there are $H$ SDT tensors at each order $m \geq m^\star$, there cannot be any other SD tensor which is not traceless but whose curls are non-zero or equivalently any RNT tensor at order $m$. Otherwise, there would be one additional SDT tensor at order $m+1$ by applying the curl operator but this would raise the number of SDT at level $m+1$ to $H+1$, which is not the case.

\item We conclude that all RNT potentials and SD tensors whose curl are non-zero are classified by the explicit tensors that we build out of terms with up to $m^\star$ derivatives. At higher order $m > m^\star$ in derivatives, all SD tensors whose curl is non-zero are traceless and can be obtained by applying curls on the RNT potentials.

\end{enumerate}

\subsection{$(\sigma,\sigma)$ SD tensors}

This analysis was already performed in \cite{Compere:2011db}. Let us summarize the results here in our new notation. We have $m^\star = 3$ because the tensor structure $\eps_a^{\; \, cd}\s_{b c}\s_{d}$ first appears at order $m=3$ and no other tensor structure appears at higher $m$. There are no tensor potentials at order $m=0$ and $m=1$. At order $m=2$, there are two independent tensor potentials
\beqs\label{M2ss}
 M^{[2,\s,\s, I]}_{ab}&=&  (5 \s^2 +\s_c \s^c ) h_{ab}^{(0)} + 4\s \s_{ab} ,\nonumber \\
 M^{[2,\s,\s,II]}_{ab} &=&  (\DD_a \DD_b + h^{(0)}_{ab}) \: \s^2\, .
 \eeqs
The first one is a RNT potential whose successive curls generate the unique SDT tensor  at each order $m \geq m^\star$, while the second one has a zero curl. The two SD tensors associated with the tensor potentials are given by
\beqs
\kappa^{[\s,\s,I]}_{ab} &=&  M^{[2,\s,\s, I]}_{ab}-  M^{[2,\s,\s, I]} h^{(0)}_{ab}  \nonumber \\
&=& (2 \s^2 -2\s_c \s^c ) h_{ab}^{(0)} + 4\s \s_{ab} \label{k2ss1} ,\\
\kappa^{[\s,\s,II]}_{ab}&=&  M^{[2,\s,\s,II]}_{ab}-  M^{[2,\s,\s,II]} h^{(0)}_{ab}  \nonumber \\
&=& 2 \s_a \s_b + 2 \s \s_{ab} + h^{(0)}_{ab} \Big( 4 \s^2 -2 \s_c \s^c \Big) \label{k2ss2} .
\eeqs
The second SD tensor has a zero symmetrized curl.

\subsection{$(k,k)$ SD tensors}

The classification of $(k,k)$ structures is considerably more involved than the classification of $(\s,\s)$ structures. In order to simplify the identification of a basis of independent tensors on-shell, we will make an efficient use of the relations
\beqs\label{ddqq}
\DD_{[a}B^{(1)}_{b]c}=0 \: , \qquad \DD_{[a}k_{b]c}=-\epsilon_{abd} \: B^{(1)\: d}_{c} \: .
\eeqs
We start by listing the independent structures quadratic in $B^{(1)}$ and its derivatives. Then, we add an independent subset of structures of the form $(B^{(1)}, k)$ such that no linear combinations are of the form $(B^{(1)}, B^{(1)})$ and eventually we add a subset of independent $(k,k)$ structures such that no linear combinations are of the form $(B^{(1)}, B^{(1)})$ or $(B^{(1)}, k)$. To look if such linear combinations exist, we just need to take into account the equations (\ref{ddqq}).

For an odd number $(2n+1)$ of derivatives we find the general form
\beqs\label{Qkkodd}
&&  Q^{(2n+1)}_{ab}= \nn\\
&& a \: \epsilon_{cd(a} \DD^{i_1}...\DD^{i_{n-1}} \DD_{b)} B^{(1)\: ce} \DD_{i_1}...\DD_{i_{n-1}} B^{(1)\:d}_{e}
 +  b \: \epsilon_{cd(a}  \DD^{i_1}...\DD^{i_{n-1}}  \DD^f \DD_{b)} k^{ce} \DD_{i_1}...\DD_{i_{n-1}}  \DD_{f} k^d_{\;e}   \nonumber \\
&& + c \:  \DD^{i_1}...\DD^{i_{n-1}}  \DD_c B^{(1)}_{d(a}  \DD_{i_1}...\DD_{i_{n-1}}  \DD^{c} k_{b)}^{\:\;d} + d \: \DD^{i_1}...\DD^{i_{n-1}}  \DD_c B^{(1)}_{de}  \DD_{i_1}...\DD_{i_{n-1}}  \DD^{c} k^{de} h^{(0)}_{ab}  \nonumber \\
 &&+ e \: \DD^{i_1}...\DD^{i_{n-1}}   \DD_c \DD_d B^{(1)}_{ab}  \DD_{i_1}...\DD_{i_{n-1}}  k^{cd}  + f\:  \DD^{i_1}...\DD^{i_{n-1}}  B^{(1)}_{cd}  \DD_{i_1}...\DD_{i_{n-1}}  \DD_{c} \DD_{(a} k_{b)}^{\:\;d} \: ,\nn
\eeqs
while for an even  number  ($2n$)  of derivatives we find
\beqs\label{Qkkeven}
  && Q^{(2n)}_{ab} =\nn \\
&& a \: \DD_{i_1}...\DD_{i_{n-2}} \DD_{c} B^{(1)}_{de} \DD^{i_1}...\DD^{i_{n-2}} \DD^c B^{(1) \: de} \: h^{(0)}_{ab} +b\:   \DD_{i_1}...\DD_{i_{n-2}} \DD_c B^{(1)}_{d(a} \DD^{i_1}...\DD^{i_{n-2}} \DD^c B^{(1) \: d}_{b)} \nonumber \\
&&+ c\:  \DD_{i_1}...\DD_{i_{n-2}} \DD_c \DD_d B^{(1)}_{ab} \DD^{i_1}...\DD^{i_{n-2}} B^{(1) \: cd} + d\: \epsilon_{cd(a} \DD^{i_1}...\DD^{i_{n-2}} \DD^f \DD^c k_{b)}^{\;\:e}  \DD_{i_1} ...\DD_{i_{n-2}} \DD_f B^{(1) \: d}_{e}  \nonumber\\
&&+ e\: \epsilon_{cd(a} \DD^{i_1}...\DD^{i_{n-2}} \DD^f  \DD_{b)} B^{(1) \: c}_{e}  \DD_{i_1} ...\DD_{i_{n-2}}  \DD_f k^{de}  +f\: \DD_{i_1}...\DD_{i_{n-2}} \DD_e \DD_{(a} k^{cd} \DD^{i_1}...\DD^{i_{n-2}} \DD^e \DD_{b)} k_{cd} \nonumber \\
 && +g\:  \DD_{i_1}...\DD_{i_{n-2}} \DD^e \DD_{(a} \DD_{b)} k^{cd} \DD^{i_1}...\DD^{i_{n-2}} \DD_e  k_{cd}
 +h\:   \DD_{i_1}...\DD_{i_{n-2}}  \DD_{c} \DD_{d} k_{ef} \DD^{i_1}...\DD^{i_{n-2}} \DD^c  \DD^d  k^{ef} h^{(0)}_{ab} \: .\nn
\eeqs
We deduce that $m^\star = 3$. Indeed, when $m=2$ the third term in $Q^{(2)}_{ab}$ does not exist while when $m=3,4$ or higher all terms in $Q^{(m)}_{ab}$ exist. Looking at $m \geq m^\star$ and imposing the SDT condition, we find after a straightforward analysis that there can be at most three independent SDT tensors. We thus have $H=3$.

We now need to construct RNT potentials and SDT tensors at each order $m \leq m^\star$. At $m=0$, there are no tensor potentials and no SDT tensors. At $m=1$, we have
\beqs
Q^{(1)}_{ab}= a \:\epsilon_{cd(a}  \DD_{b)} k^{ce}  k^{d}_{\;\;e} + b \: B^{(1)}_{c(a} k_{b)}^{\:\;c} + c \:  B^{(1)}_{cd} k^{cd} h^{(0)}_{ab} \: ,
\eeqs
and one easily checks that there are no SDT tensors but there is one RNT potential :
\beqs
M^{[1,k,k]}_{ab}= -4 B^{(1)}_{c(a} k_{b)}^{\:\:c} +B^{(1)}_{cd} k^{cd} h^{(0)}_{ab}.
\eeqs
At $m=2$, we have:
\beqs
Q^{(2)}_{ab}&=& a \:   B^{(1)}_{cd}  B^{(1) \: cd} \: h^{(0)}_{ab} +b\:  B^{(1)}_{c(a}   B^{(1) \: c}_{b)}  + c\: \epsilon_{cd(a}  \DD^c k_{b)}^{\;\:e}  B^{(1) \: d}_{e}  + d\: \epsilon_{cd(a}   \DD_{b)} B^{(1) \: c}_{e}   k^{de}  \nonumber \\
 && +e\:  \DD_{(a} k^{cd} \DD_{b)} k_{cd}  +f\:  k^{cd}  \DD_{(a} \DD_{b)}    k_{cd}   +g\:   \DD_{c} k_{de}  \DD^c   k^{de} h^{(0)}_{ab}  \nonumber \\
 && + h\: k^c_{\: (a} k_{b)c} + i \: k_{cd} k^{cd} h^{(0)}_{ab} \:,
\eeqs
where we also introduced the  structures with $m=0$ derivatives. Here, we get:
\beqs
Q^{(2)}&=& [3a+b-2c] B^{(1)\:cd} B^{(1)}_{cd} + \DD_b k_{cd} \DD^b k^{cd} [e+3g] +[3f+h+3i] k_{cd} k^{cd} \: ,\nonumber \\
\DD_a Q^{(2)}&=&[6a+2b-4c] B^{(1)\:cd} \DD_a B^{(1)}_{cd}+ [2e+6g] \DD_a \DD_b k_{cd} \DD^b k^{cd} \nonumber \\
&&  + [6f+2h+6i] k_{cd} \DD_a k^{cd} \: , \nonumber \\
\DD^b Q^{(2)}_{ab}&=& [ 2a+b-2d] B^{(1)\:cd} \DD_a B^{(1)}_{cd} + [\frac{c}{2} -\frac{d}{2}] \epsilon_{cda}\DD^c k_{b}^{\:\:e} \DD^b B^{(1)\:d}_e \nonumber \\
&&+ [2c-2d-4e+4f+2h] \:  \epsilon_{cda} k^{ce} B^{(1)\:d}_e  +[e+f+2g]  \DD_a \DD^b k_{cd} \DD_b k_{cd} \nonumber\\
&&+k^{cd} \DD_a k_{cd} [e+7f+h+2i] \: ,
\eeqs
where we made use of the relations:
\beqs
&& k^{cd} \square \DD_a k_{cd} = 5 k^{cd} \DD_a k_{cd} + 4 k^{cd} \DD_c k_{ad} , \qquad k^{cd} \DD^b \DD_c \DD_d k_{ab}=7 k^{cd} \DD_c k_{ad}  \: , \nonumber \\
&& k^{cd} \DD^b \DD_a \DD_b k_{cd}= k^{cd} \square \DD_a k_{cd}- 2k^{cd} \DD_c k_{ad} ,\qquad  k^{cd} \DD_c k_{ad}=2 \epsilon_{cda} k^c_{\:\:e} B^{(1)\:d}_{e}+k^{cd} \DD_a k_{cd}\, .\nn
\eeqs

We obtain that tensors $Q^{(2)}_{ab}$ satisfying $\DD^b Q^{(2)}_{ab}=\DD_a Q^{(2)}$ are of the form
\beqs
 m_1 Y^{(2)}_{ab} + m_2 M^{[2,k,k,I]}_{ab}+ m_3 M^{[2,k,k,II]}_{ab} +m_4 M^{[2,k,k,III]}_{ab} \: ,
\eeqs
where
\beqs
Y^{(2)}_{ab} &=& -4 \:  B^{(1)}_{c(a}   B^{(1) \: c}_{b)}  -2\: \epsilon_{cd(a}  \DD^c k_{b)}^{\;\:e}  B^{(1) \: d}_{e}  -2\: \epsilon_{cd(a}   \DD_{b)} B^{(1) \: c}_{e}   k^{de}= \epsilon_{cd(a} \DD^c M^{(1)\:d}_{b)} \: , \label{Y2aa}\nonumber \\
M^{[2,k,k,I]}_{ab}&=& \frac{1}{8}k_{cd} k^{cd} h_{ab}^{(0)} - k_{ac}k^c_{\; b} + \frac{1}{8}\DD_c k_{de}\DD^c k^{de} h_{ab}^{(0)}-\frac{1}{2}\DD_a k_{cd}\DD_b k^{cd} \: , \nonumber \\
M^{[2,k,k,II]}_{ab}&=&  -\frac{1}{4}B^{(1)}_{cd}  B^{(1) \: cd} \: h^{(0)}_{ab} +\:  B^{(1)}_{c(a}   B^{(1) \: c}_{b)}  \: ,\nonumber \\
M^{[2,k,k,III]}_{ab}&=& \frac{1}{8} k^{cd} \DD_{(a} \DD_{b)} k_{cd} +\frac{1}{8} \DD_{(a} k^{cd} \DD_{b)} k_{cd} +  \frac{1}{16} k_{cd} k^{cd}h^{(0)}_{ab} \: .
\eeqs
We thus see that $Y^{(2)}_{ab}$  is the unique SDT tensor obtained from the RNT potential $M^{[1,k,k]}_{ab}$, that $M^{[2,k,k,I]}_{ab}$ and $M^{[2,k,k,II]}_{ab}$ are two new RNT potentials and that $M^{[2,k,k,III]}_{ab}$ is a tensor potential which has vanishing curl as it is of the form $(\DD_a \DD_b +h^{(0)}_{ab})k_{cd} k^{cd}$.
From the three tensor potentials found, we can define 3 SD tensors:
\beqs
\kappa^{[k,k,I]}_{ab} &=& (M^{[2,k,k,I]}_{ab}- M^{[2,k,k,I]} h^{(0)}_{ab})  \nonumber \\
        &=& \frac{3}{4} k_{cd} k^{cd} h_{ab}^{(0)} -  k_{ac}k^c_{\; b} + \frac{1}{4} \DD_c k_{de}\DD^c k^{de} h_{ab}^{(0)}-\frac{1}{2} \DD_a k_{cd}\DD_b k^{cd} \: , \nonumber \\
\kappa^{[k,k,II]}_{ab}&=& (M^{[2,k,k,II]}_{ab}- M^{[2,k,k,II]} h^{(0)}_{ab}) = -\frac{1}{2} B^{(1)\: cd} B^{(1)}_{cd} h^{(0)}_{ab} +  B^{(1)}_{c(a} B^{(1)\: c}_{b)} \: , \nonumber \\
\kappa^{[k,k,III]}_{ab} &=&  (M^{[2,k,k,III]}_{ab} -M^{[2,k,k,III]} h^{(0)}_{ab}) \nonumber \\
                             &=& (-\frac{1}{8} \DD_c k_{de} \DD^c k^{de}  -\frac{1}{2} k_{cd} k^{cd} ) h^{(0)}_{ab}+\frac{1}{8} \DD_{(a} k^{cd} \DD_{b)} k_{cd} +\frac{1}{8} k^{cd} \DD_{(a} \DD_{b)} k_{cd} \: .
\eeqs

Now, at $m=3$, we obtain:
\beqs
Q^{(3)}_{ab}&=& a \: \epsilon_{cd(a}  \DD_{b)} B^{(1)\: ce}  B^{(1)\:d}_{e}  +  b \: \epsilon_{cd(a}   \DD^f \DD_{b)} k^{ce}   \DD_{f} k^d_{\;e}    + c \:   \DD_c B^{(1)}_{d(a}   \DD^{c} k_{b)}^{\:\;d} \nonumber \\
&& + d \:  \DD_c B^{(1)}_{de}  \DD^{c} k^{de} h^{(0)}_{ab} + e \:    \DD_c \DD_d B^{(1)}_{ab}    k^{cd}  + f\:   B^{(1)}_{cd}   \DD_{c} \DD_{(a} k_{b)}^{\:\;d} \: ,
\eeqs
up to terms with lower derivatives. One can explicitly construct three independent SDT tensors which can also be obtained as curls of the three previous RNT potentials. We have thus completed our algorithm. We have three towers of SDT tensors generated by the three RNT potentials $M^{[1,k,k]}_{ab}$  (which leads to $Y^{(2)}_{ab}$), $M^{[2,k,k,I]}_{ab}$ and $M^{[2,k,k,II]}_{ab}$.

\subsection{$(\sigma,k)$ SD tensors}

For the $(\sigma,k)$ case, we find that a generic tensor with $2n+1$ derivatives is of the form
\beqs
Q^{(2n+1)}_{ab}&=& a \; \DD_{i_1}...\DD_{i_{n}} B^{(1)}_{ab} \sigma^{i_1...i_{n}} + b \; \DD_{i_1}...\DD_{i_{n-1}} B^{(1)}_{e(a} \sigma_{b)}^{\;\:e i_1 ...i _{n-1}}  \nonumber \\
&&+ c \; \epsilon_{cd(a} \DD^{i_1} ...\DD^{i_{n-1}} \DD_{b)} k^c_{\;e} \sigma^{de}_{\;\;\;i_1 ... i_{n-1}} +  d\; \epsilon_{cd(a} \DD^{i_1} ...\DD^{i_{n-1}} k^{ce} \sigma_{b)\;\;\;e i_{1} ...i_{n-1}}^{\;\;d}\nonumber \\
&& + e \; \DD_{i_1}...\DD_{i_{n-1}} B^{(1)}_{cd} \sigma^{cd i_1...i_{n-1}} h^{(0)}_{ab}  + f \; \DD_{i_1}...\DD_{i_{n-2}} B^{(1)}_{cd} \sigma_{ab}^{\;\;\;cd i_1...i_{n-2}} \: ,
\eeqs
while, for an even number $2n$ of derivatives, it is of the form
\beqs
Q^{(2n)}_{ab}&=& a \; \sigma^{c d i_1...i_{n-2}} \DD_{i_1}...\DD_{i_{n-2}} \DD_c \DD_{(a}k_{b)d}    +\: b \; \sigma^{c d  i_1...i_{n-2}} \DD_{i_1}...\DD_{i_{n-2}} \DD_{(a} \DD_{b)} k_{cd} \nonumber \\
             &&   +\: c \; \sigma^{ cd  i_1...i_{n-2}} \DD_{i_1}...\DD_{i_{n-2}} \DD_c \DD_d  k_{ab} + \:  d \; \sigma_{c d i_1...i_{n-2} (a} \DD^{i_1}...\DD^{i_{n-2}} \DD_{b)}k^{cd} \nonumber \\
             && +\:  e \;  \sigma_{c d  i_1...i_{n-2} (a} \DD^{i_1}...\DD^{i_{n-2}} \DD^c k_{b)}^{\:\:\:d} + \: f \; \sigma_{a b c d  i_1...i_{n-2}} \DD^{i_1}...\DD^{i_{n-2}} k^{cd} \nonumber \\
             && + \: g \; h^{(0)}_{ab} \: \sigma_{c d  i_1...i_{n-1}} \DD^{i_1}...\DD^{i_{n-1}} k^{cd} \: .
\eeqs
We find that $m^\star =4$. Indeed, for $m=3$ derivatives, the last term in $Q^{(3)}_{ab}$ does not exist while for $m=4,5,\dots$ all terms in $Q^{(m)}_{ab}$ exist. For any $m \geq 4$, one can check that there are at most 2 SDT tensors. We therefore find $H=2$. At lower levels $m=0$ or $m=1$, we see that there are no SDT tensors and no tensor potentials. At $m=2$, we have
\beqs
Q^{(2)}_{ab}&=& a \:  \sigma k_{ab} +
b \: \epsilon_{cd(a} k^d_{\:\: b)} \sigma^c+ c \:   \sigma B^{(1)}_{ab}
+ d \:  \sigma^c \DD_c k_{ab}  \nonumber \\
&&+ e \:  \sigma^c \DD_{(a} k_{b)c} + f \:  \sigma_{c(a} k_{b)}^{\:\:\:c} + g \:  \sigma_{cd} k^{cd} h^{(0)}_{ab}\label{2der} \: .
\eeqs
There is no SDT tensor, but there is one RNT potential
\bea
M_{ab}^{[2,\s,k]} =   \sigma k_{ab} - \:  \sigma^c \DD_c k_{ab} +  \:  \sigma^c \DD_{(a} k_{b)c} +  \:  \sigma_{c(a} k_{b)}^{\:\:\:c} -\frac{1}{4} \:  \sigma_{cd} k^{cd} h^{(0)}_{ab}  \label{Mprimitive2} \, .
\eea
Its curl gives an SDT tensor
\beqs
Z^{(3)}_{ab}&=& 8 \sigma B^{(1)}_{ab} +\frac{1}{2} \epsilon_{cd(a} \sigma_{b)}^{\;\:ce} k^d_{\;e} + 5 \sigma^c_{(a} B^{(1)}_{b)c}  \nonumber \\
&&-\frac{1}{2} \epsilon_{cd(a} \sigma^{ce} \DD_{b)} k^d_{\;e} - 2 h^{(0)}_{ab} \sigma^{cd} B^{(1)}_{cd} -\sigma^{c} \DD_c B^{(1)}_{ab} \: .
\eeqs
At $m=3$, the general SDT tensor or RNT potential can be written as a linear combination of a basis of terms with 1 and 3 derivatives (terms with 2 derivatives decouple)
\beqs
Q^{(3)}_{ab}&=&  a \: B^{(1)e}_{(a}\sigma_{b)}^{\;\:e} + b \: h^{(0)}_{ab} \sigma^{cd} B^{(1)}_{cd} + c \: \eps_{cd(a }\s_{b)}^{\; c e}k^d_{\; e}  + d \:  \eps_{cd(a}\DD_{b)} k_{\; e}^{d}\s^{ce}+ e \: \sigma^e \DD_e B^{(1)}_{ab} \nonumber \\
&& + f \epsilon_{cd(a} k^d_{\;b)} \sigma^c + g \: \sigma B^{(1)}_{ab}\, .
\eeqs
We find one SDT tensor which is obviously $Z^{(3)}_{ab}$ and a new RNT potential $M^{[3,\s,k]}_{ab}$
\beqs
 M^{[3,\s,k]}_{ab}&=& 7 B^{(1)}_{c(a} \sigma^c_{b)}  - \frac{5}{2} h^{(0)}_{ab} \sigma^{cd} B^{(1)}_{cd} +\frac{1}{2} \epsilon_{cd(a} \sigma_{b)}^{\;\:ce} k^d_{\;e} \nn \\
 &&  -\frac{1}{2} \epsilon_{cd(a}  \DD_{b)} k^d_{\;e} \sigma^{ce} -\sigma^{c} \DD_c B^{(1)}_{ab}+ 10 \sigma B^{(1)}_{ab}\, .
\eeqs
 As expected, at $m=4$, one can check that we have 2 SDT tensors. The algorithm is therefore completed. The two RNT potentials that generate the two independent towers of SDT tensors are $M^{[2,\s,k]}_{ab}$ and $M^{[3,\s,k]}_{ab}$.
To each of these potentials corresponds a unique SD tensor
\bea
\kappa_{ab}^{[\s,k, I]} &=& M_{ab}^{[2,\s,k]}- h_{ab}^{(0)} M^{[2,\s,k]} \nonumber \\
&=& \s k_{ab} +\s_{c(a} k_{b)}^{\:\:\:c}  -\frac{1}{2} \s_{cd} k^{cd} h^{(0)}_{ab} -\s^c \DD_c k_{ab} + \s^c \DD_{(a} k_{b)c} \: , \label{k2sk}\\
\kappa_{ab}^{[\s,k,II]} &=& M_{ab}^{[3,\s,k]}- h_{ab}^{(0)} M^{[3,\s,k]} \nonumber \\
&=& Z^{(3)}_{ab} + 2 B^{(1)}_{c(a} \s_{b)}^{\:\:\:c} -\s^{cd} B^{(1)}_{cd} \: h^{(0)}_{ab} +2 \s B^{(1)}_{ab} \label{k3sk}
 \, .
\eea
All other SDT tensors, with $m \geq m^{*}$, are then constructed from linear combinations of successive curls of those SD tensors. This ends the classification of RNT and SDT $(\s,k)$ tensors.

\subsection{Properties}
\label{kappaprop}

In order to simplify the linearization stability constraints given in Section \ref{sec:LS}, we would like to understand what currents $\kappa_{ab} \: \xi^{a}_{(0)}$ associated with SD tensors $\kappa_{ab}$ constructed in the previous subsections can be expressed as total divergences. Here, we show that the currents associated with the two independent $(\s,k)$ SD tensors and two linear combinations of the three $(k,k)$ SD tensors which are not traceless are total divergences. We are thus left with two currents, one quadratic in $\s$ and one in $k$, which can be identified as the currents constructed using the stress-energy tensors of the counter-term actions $S^{(\s)}$ and $S^{(k)}$ up to total divergences.

For the $(\s,k)$ SD tensors, one can first check that the current:
\beqs\label{k1sk2}
\kappa^{[\s,k,I]}_{ab} \xi^b =\DD^b \Big( -\xi_{[a} \: k_{b]}^{\:\:\:c} \s_c +\DD^c \xi_{[a} \: \s k_{b]c} +\xi^c \: \s \DD_{[a} k_{b]c} +\xi^c \: \s_{[a} k_{b]c}   \Big)
\eeqs
can be expressed as a total divergence and will thus not contribute to the charges. As we know from \cite{Compere:2011db}, this also implies that its symmetrized curl, the SDT tensor  $Z^{(3)}_{ab}$, will not contribute to the charges. If we look at the definition of $\kappa^{[\s,k,II]}_{ab}$ given in (\ref{k3sk}), we are left with checking that
\beqs
\Big(  2 B^{(1)}_{c(a} \s_{b)}^{\:\:\:c} -\s^{cd} B^{(1)}_{cd} \: h^{(0)}_{ab} +2 \s B^{(1)}_{ab} \Big) \xi^b = 2    \DD^b \Big( \DD^c \xi_{[a} \: \s B^{(1)}_{b]c} -\xi_{[a} \: B^{(1) \: c}_{b]}  \s_c +\xi^c \: \s_{[a} B^{(1)}_{b]c}   \Big)
\eeqs
to see that $\kappa^{[\s,k,II]}_{ab}$ will not contribute either to the charges. This proves that no SDT or SD tensors quadratic in $(\s,k)$ will contribute to the charges. For the terms quadratic in $(k,k)$, one can show by inspection that:
\beqs\label{int1}
&&\DD^b \Big( 2 \xi_c \: k_{d[a} \DD^d k_{b]}^{\:\:\:c} -\DD_c \xi_d \: k^{c}_{\:\:[a} k_{b]}^{\:\:\:d} \Big) \nonumber \\
&&\qquad = \Big( \DD_c k_{d(a} \DD^d k_{b)}^{c}+4 k_{c(a} k_{b)}^{\:\:\:c}-k^{cd} \DD_c \DD_d k_{ab} \Big) \xi^b \nonumber \\
&&\qquad = \Big( 7 \:   B^{(1)}_{cd}  B^{(1) \: cd} \: h^{(0)}_{ab} - 6\:  B^{(1)}_{c(a}   B^{(1) \: c}_{b)}  + 4\: \epsilon_{cd(a}  \DD^c k_{b)}^{\;\:e}  B^{(1) \: d}_{e}  + 4\: \epsilon_{cd(a}   \DD_{b)} B^{(1) \: c}_{e}   k^{de}  \nonumber \\
 && \qquad \qquad   + \:  \DD_{(a} k^{cd} \DD_{b)} k_{cd}  - \:  k^{cd}  \DD_{(a} \DD_{b)}    k_{cd}   + 4 \: k^c_{\: (a} k_{b)c} +  \: k_{cd} k^{cd} h^{(0)}_{ab}  \Big) \xi^b \nonumber \\
 &&\qquad = \Big( - 2 Y^{(2)}_{ab}- 4 \kappa^{[k,k,I]}_{ab} - 14  \kappa^{[k,k,II]}_{ab} -8 \kappa^{[k,k,III]}_{ab}\Big)  \xi^b \: ,
\eeqs
and also:
\beqs\label{int2}
&&\DD^b \Big( \DD_c \xi_{[a} \: k_{b]}^{\:\:\:d} k^c_{\:\:d} -\xi_{[a} \: \DD^c k_{b]}^{\:\:\:d} k_{cd} -\xi^c  \: k^d_{\:\:[a} \DD_{b]} k_{cd} +\xi^c \: k_{cd} \DD_{[a} k_{b]}^{\:\:\:d}\Big) \nonumber \\
&&\qquad = \Big( -2  \: k^c_{\: (a} k_{b)c} -\frac{3}{2}  \: k_{cd} k^{cd} h^{(0)}_{ab} -\frac{1}{2} \DD_c k_{de} \DD^d k^{ce} h^{(0)}_{ab} \nonumber \\
 && \qquad \qquad + \DD^c k_{(a}^{d} \DD_{b)} k_{cd} -\DD_c k_{d(a} \DD^c k_{b)}^{d}+k^{cd} \DD_c \DD_{(a} k_{b)d} \Big) \xi^b \nonumber \\
&&\qquad = \Big(-5 \:   B^{(1)}_{cd}  B^{(1) \: cd} \: h^{(0)}_{ab} +6 \:  B^{(1)}_{c(a}   B^{(1) \: c}_{b)}  -2 \: \epsilon_{cd(a}  \DD^c k_{b)}^{\;\:e}  B^{(1) \: d}_{e}  -2 \: \epsilon_{cd(a}   \DD_{b)} B^{(1) \: c}_{e}   k^{de}  \nonumber \\
 && \qquad \qquad  + \:  k^{cd}  \DD_{(a} \DD_{b)}    k_{cd}   -\frac{1}{2}\:   \DD_{c} k_{de}  \DD^c   k^{de} h^{(0)}_{ab}   -2\: k^c_{\: (a} k_{b)c} -\frac{5}{2} \: k_{cd} k^{cd} h^{(0)}_{ab}  \Big) \xi^b \nonumber \\
&&\qquad = \Big( Y^{(2)}_{ab}+ 2 \kappa^{[k,k,I]}_{ab}+ 10 \kappa^{[k,k,II]}_{ab} + 8 \kappa^{[k,k,III]}_{ab} \Big)  \xi^b \: .
\eeqs
The SD tensor $\kappa^{[k,k,III]}_{ab}$ has a trivial symmetrized curl. It is thus associated with trivial charges when contracted with a Killing vector. This result is also reflected in the fact that the current
\beqs\label{totdiv}
\kappa^{[k,k,III]}_{ab} \xi^b=-\frac{1}{8} \DD^b \Big( 2 k^{cd} \xi_{[a}  \DD_{b]} k_{cd} +\frac{1}{2} k_{cd} k^{cd} \DD_{[a} \xi_{b]}  \Big)
\eeqs
can be written as a total divergence. Eventually, we notice from (\ref{int1}) and (\ref{int2}), that the current constructed out of $\kappa^{[k,k,II]}_{ab}$ can be written as a total divergence and that the equality
\beqs
Y^{(2)}_{ab} \: \xi^{(0)\:b}=-2 \kappa^{[k,k,I]}_{ab} \: \xi^{(0)\:b}
\eeqs
is true up to a total divergence. We have thus shown  in particular that the charges constructed from the current $\kappa_{ab} \: \xi^a_{(0)}$ where $\kappa_{ab}$ is defined in \eqref{defkappa} can be written as
\beqs
\oint d^2 S \sqrt{-h^{(0)}} \kappa_{ab} \: \xi^a_{(0)} \: n^b = \oint d^2 S \sqrt{-h^{(0)}} \: (  \kappa_{ab}^{[\s,\s,I]}+ \frac{1}{2}  Y^{(2)}_{ab})\: \xi^a_{(0)} \: n^b\, .
\eeqs

\section{Equations of motion in Beig-Schmidt form}
\label{appEOM}

In this appendix, we will first review the 3+1 split of Einstein's equations $G_{\mu\nu}=0$ for the four-dimensional metric $g_{\mu\nu}$ on three-dimensional hypersurfaces $h_{ab}$ which have a spacelike normal $n^\mu$, $n^\mu n_\mu = +1$. We will then expand those equations asymptotically using Beig-Schmidt coordinates up to second order in the expansion while keeping $\s \neq 0$, $k_{ab} \neq 0$ and $i_{ab} \neq 0$. This extends all previous treatments, see e.g. \cite{BS,B,Beig:1987aa,deHaro:2000wj,Mann:2005yr,Mann:2006bd,Mann:2008ay}. We will eventually rewrite the second order equations in terms of two, much more compact, equivalent systems.

Our notation is as follows. We use greek letters $\mu,\nu,...$ for four-dimensional quantities and latin letters $a,b,...$ for three-dimensional quantities. The letter $\rho$ always denotes the radial coordinate that we normalize as $n^\mu \partial_\mu= (1+\frac{\sigma}{\rho})^{-1}\frac{\partial}{\partial \rho}$. We define the metric $h_{\mu\nu} = g_{\mu\nu} -n_{\mu} n_{\nu} $. The extrinsic curvature is defined by $K_{ab} \equiv h_{a}^{\: \mu} h_{b}^{\: \nu} \nabla_{\mu} n_{\nu} $ and its trace is $K=h^{ab} K_{ab}$. In our case, it reduces to $K_{ab} = \frac{1}{2}n^\rho \partial_\rho h_{ab}$.

Covariant derivatives associated  with  $g_{\mu\nu}$, $h_{ab}$, and $h^{(0)}_{ab}$ are respectively denoted by $\nabla_{\mu}$, $D_a$ and $\DD_{a}$.  We have $D_{a}=h_{a}^{\:\:\mu} \nabla_{\mu}$. The four-dimensional curvature, Ricci tensor and scalar are denoted by $R^{\mu}_{\:\: \nu\rho\sigma}$, $R_{\mu\nu}$ and $R$ while their 3d counterparts are denoted by $\mathcal{R}^a_{\:\:bcd}$,$\mathcal{R}_{ab}$ and $\mathcal{R}$. We use the Misner-Thorne-Wheeler conventions $R^a_{\; \, bcd} =R_{cd\;\; b}^{\;\;\; a}= \partial_c \Gamma^a_{bd} + \Gamma^a_{ce}\Gamma^e_{bd} - (c \leftrightarrow d)$, $R_{ab} = R^c_{\;\, acb}$.

\subsection{The 3+1 split}
\label{eq:31split}

Einstein's equations can be split into a set of three equations when appropriately projected along perpendicular and/or normal directions to the hypersurface.
This  provides us with Hamiltonian and momentum equations of motion (these equations contain time derivatives and therefore are not constraints) and equations of motion on the 3-dimensional hypersurface which respectively read
\beqs \label{3equations}
H &\equiv& -2 G_{\mu\nu} n^{\mu} n^{\nu}= \mathcal{R} +K_{ab} K^{ab} -K^2=0  \: , \nonumber \\
F_{a} &\equiv& h_{a}^{\:\:\mu} n^{\nu} R_{\mu\nu} =D_{b} K^{b}_{\:\:a}-D_a K=0  \: , \\
F_{ab} &\equiv&   h_{a}^{\:\:\mu} h_{b}^{\:\:\nu} R_{\mu\nu}
              = \mathcal{R}_{ab}-(\mathcal{L}_{n}K)_{ab}+D_{(a} a_{b)}-a_{a}a_{b}-K K_{ab}+ 2 K_{a}^{\:\:c} K_{cb}=0  \: , \nonumber
\eeqs
where $\mathcal{L}_{n}$ stands for the Lie derivative along the unit normal and $a_\mu \equiv n^\rho \nabla_{\rho} n_\mu$ is the 4-acceleration.

The Hamiltonian equation can be further simplified by taking the trace of the equation of motion so that we get
\begin{eqnarray}
H\equiv -\mathcal{L}_{n} K -K_{ab}K^{ab}+D_a a^{a}-a_a a^{a}=0 \: .
\end{eqnarray}
Using our values for the shift $N^a=0$, we get $D_{(a} a_{b)}-a_{a}a_{b} = -N^{-1} D_a D_b N$ and the equations reduce to the ones of Beig and Schmidt \cite{BS}
\beqs\label{3equationssimp}
H &\equiv&  -\mathcal{L}_{n} K - K_{ab}K^{ab}-N^{-1} h^{ab} D_a D_b N=0  \: , \nonumber \\
F_{a} &\equiv& D_{b} K^{b}_{\:\:a}-D_a K=0  \: , \nonumber \\
F_{ab} &\equiv &   \mathcal{R}_{ab}- N^{-1} \partial_{\rho} K_{ab}-N^{-1} D_a D_b N-K K_{ab}+ 2 K_{a}^{\:\:c} K_{cb}=0  \: .
\eeqs

\subsection{Equations of motion in the radial expansion}

We now expand these equations using our boundary conditions \eqref{metric1}. The inverse metric is expanded as
\begin{eqnarray*}
h^{ab}(\rho,x^c)&=& \rho^{-2} h^{(0) ab}-\rho^{-3} h^{(1) ab}-\ln\rho\, \rho^{-4} i^{ab}-\rho^{-4} (h^{(2) ab}-h^{(1)a}_{\quad \: c} h^{(1)cb})+O(\rho^{-5}) \; . \nonumber
\end{eqnarray*}
The extrinsic curvature admits the simple expansion
\bea
K_{ab} = \rho \, h^{(0)}_{ab} + \left( \frac{1}{2} h^{(1)}_{ab} -\s h^{(0)}_{ab} \right) + \frac{1}{\rho} \left( \frac{1}{2}i_{ab} - \frac{1}{2} h^{(1)}_{ab} \s + \s^2 h^{(0)}_{ab}   \right)+O(\rho^{-2}) \, ,
\eea
and we also have
\bea
K^a_{\; b} &=& \frac{1}{\rho} \delta^a_b - \frac{1}{2\rho^2} k^a_{\; b}-\frac{\ln\rho}{\rho^3} i^a_{\; b} +\frac{1}{\rho^3} \left( -h_{(2)\; b}^a+\frac{1}{2}i^a_{\; b} +2\s^2 \delta^a_b + \frac{1}{2} k^a_{\; c}k^c_{\; b} - \frac{3}{2} \sigma k^a_{\; b}  \right) +O(\rho^{-4}) \; .\nn
\eea
The covariant derivative requires an expansion of the Christoffel symbols
\begin{eqnarray}
\Gamma^a_{bc} = \Gamma^{(0)\:a}_{\quad\:\: bc}+ \rho^{-1} \Gamma^{(1)\:a}_{\quad\:\: bc}+\ln\rho\,  \rho^{-2}\Gamma^{(ln,2)\:a}_{\quad\:\: bc}+ \rho^{-2}\Gamma^{(2)\:a}_{\quad\:\: bc}+ O(\rho^{-3})  \: ,
\end{eqnarray}
where
\bea
\Gamma^{(1)\:a}_{\quad\:\: bc} &=& \frac{1}{2}\left( \DD_c h^{(1)a}_{\;\; b} +  \DD_b h^{(1)a}_{\;\; c} -  \DD^a  h^{(1)}_{bc}\right)  \: , \nn\\
\Gamma^{(ln,2)\:a}_{\quad\:\: bc} &=& \frac{1}{2}\left( \DD_c i^{a}_{\;\; b} +  \DD_b i^{a}_{\;\; c} -  \DD^a  i_{bc}\right)\, ,\\
\Gamma^{(2)\:a}_{\quad\:\: bc} &=& \frac{1}{2}\left( \DD_c h^{(2)a}_{\;\; b} +  \DD_b h^{(2)a}_{\;\; c} -  \DD^a  h^{(2)}_{bc}\right) - \frac{1}{2}h^{(1) ad}
\left( \DD_c h^{(1)}_{db} +  \DD_b h^{(1)}_{dc} -  \DD_d  h^{(1)}_{bc}\right)   \: .\nn
\eea
The expansion of the three-dimensional Ricci curvature tensor is
\begin{eqnarray}\label{riccitensor}
\mathcal{R}_{ab}= \mathcal{R}^{(0)}_{ab}+ \rho^{-1}\mathcal{R}^{(1)}_{ab}+\ln\rho  \rho^{-2}\mathcal{R}^{(ln,2)}_{ab}+ \rho^{-2}\mathcal{R}^{(2)}_{ab}  +O(\rho^{-3})  \: .
\end{eqnarray}
The zeroth order Ricci tensor is the one constructed with the metric $h^{(0)}_{ab}$. The first order Ricci tensor and the tensor $\mathcal{R}^{(ln,2)}_{\:\:\:\:ab}$ are

\begin{eqnarray*}
\mathcal{R}^{(1)}_{\:\:\:\:ab}&=& \DD_c \biggr [ \Gamma^{(1)\:c}_{\quad \:\:ab} \biggl ]-\DD_b \biggr [ \Gamma^{(1)\:c}_{\quad \:\:ac} \biggl ]
= \frac{1}{2} \biggr [ \DD^{c} \DD_{b} h^{(1)}_{ac}+ \DD^{c} \DD_{a} h^{(1)}_{bc} -\DD_{a} \DD_{b} h^{(1)} - \DD^{c} \DD_{c} h^{(1)}_{ab} \biggl ]   \: ,\\
\mathcal{R}^{(ln,2)}_{\:\:\:\:ab} &=&\frac{1}{2} \biggr [ \DD^{c} \DD_{b} i_{ac}+ \DD^{c} \DD_{a} i_{bc} -\DD_{a} \DD_{b} i - \DD^{c} \DD_{c} i_{ab} \biggr ] ,
\end{eqnarray*}
and the second order Ricci tensor reads as
\begin{eqnarray*}
\mathcal{R}^{(2)}_{\:\: ab} &=& \frac{1}{2} \biggr [ \DD^{c} \DD_{b} h^{(2)}_{ac}+ \DD^{c} \DD_{a} h^{(2)}_{bc} -\DD_{a} \DD_{b} h^{(2)} - \DD^{c} \DD_{c} h^{(2)}_{ab} \biggl ] +\frac{1}{2} \DD_b \biggr [ h^{(1)cd} \DD_a h^{(1)}_{cd} \biggl ]\nonumber \\
&& -\frac{1}{2} \DD_d \biggr [ h^{(1)cd} (\DD_a h^{(1)}_{bc}+\DD_b h^{(1)}_{ac}-\DD_c h^{(1)}_{ab} ) \biggl ]  +\frac{1}{4} \DD^{c} h^{(1)} \biggr [ \DD_a h^{(1)}_{bc}+\DD_b h^{(1)}_{ac}-\DD_c h^{(1)}_{ab}\biggl ]\nonumber \\
&&-\frac{1}{4} \DD_{a} h^{(1)}_{cd} \DD_{b} h^{(1)cd}  +\frac{1}{2} \DD_{c} h^{(1)}_{ad} \DD^{c} h^{(1)d}_{\:\:\:\:\:\: \:\:b} -\frac{1}{2} \DD_{c} h^{(1)}_{ad} \DD^{d} h^{(1)c}_{\:\:\:\:\:\: \:\:b}\, .
\end{eqnarray*}
Finally, the equations can be expanded as
\begin{eqnarray}
H &=& \rho^{-3} H^{(1)} +\ln\rho\, \rho^{-4} H^{(ln,2)} +\rho^{-4} H^{(2)} +O(\rho^{-5})  \: , \label{equation1bis} \nonumber \\
F_a &=& \rho^{-2} F^{(1)}_a+\ln \rho\, \rho^{-3} F^{(ln,2)}_a+\rho^{-3} F^{(2)}_a+O(\rho^{-4})   \: , \label{equation2}  \label{equation3}\\
F_{ab}&=& F^{(0)}_{ab} + \rho^{-1} F^{(1)}_{ab}+ \ln\rho \, \rho^{-2} F^{(ln,2)}_{ab}+\rho^{-2} F^{(2)}_{ab} +O(\rho^{-3})  \: .\nonumber
\end{eqnarray}
At zeroth order, we only have $F^{(0)}_{ab}=\mathcal{R}^{(0)}_{ab}-2 h^{(0)}_{ab}=0 $ which implies that the boundary metric is three-dimensional de Sitter spacetime.
At first order, the Hamiltonian equation $H^{(1)}=0$ is simply
\bea
\fbox{
$(\Box +3 )\sigma = 0$
}
\eea
The momentum equation $F^{(1)}_a = 0$ is
\beqs\label{divk1}
\fbox{$ \DD^b k_{ab}=0$}
\eeqs
and the radial equation of motion $F^{(1)}_{ab}$ is
\bea\fbox{
$(\Box - 3 )k_{ab} = 0$}
\eea
after we set $k=0$.
At second order we easily get for the logarithmic terms $ H^{(ln,2)} = 0$, $F^{(ln,2)}_a = 0$ and $ F^{(ln,2)}_{ab} = 0$
\begin{eqnarray}\label{eqh2a}
\fbox{
   $i = 0,\qquad \DD^b i_{ab}=0, \qquad (\Box-2) i_{ab}=0$
}
\end{eqnarray}
For the finite terms at second order we find
\begin{eqnarray}
H^{(2)}&=& -h^{(2)}+\frac{3}{2} i +\frac{1}{4}h^{(1)ab} h^{(1)}_{ab}+  \frac{1}{2} \sigma h^{(1)} \nonumber \\
&& +9 \sigma^{2}+\sigma \DD^2 \sigma+ h^{(1)ab} \DD_a \DD_b \sigma +  \DD^{b} \sigma \DD^{a} h^{(1)}_{ab}  -\frac{1}{2} \DD_a \sigma \DD^{a} h^{(1)} \: .
\end{eqnarray}
Using only $\sigma$ and $k_{ab} = h_{ab}^{(1)} + 2\s h_{ab}^{(0)}$, and also $k=i=0$, we obtain
\begin{eqnarray}\label{eqh2b}
\fbox{
   $ h^{(2)}=12 \sigma^2 + \sigma_c \sigma^c+\frac{1}{4} k_{cd} k^{cd}  + k_{cd} \sigma^{cd}$
}
\end{eqnarray}
where we also made use of the first order equations of motion. We also have:
\be
F^{(2)}_a \equiv \DD_b K^{(2)b}_a - \DD_a K^{(2)} + \Gamma_{bc}^{(1)b}K^{(1)c}_{\; a} - \Gamma_{ab}^{(1)c}K^{(1)b}_{c}=0 \: ,
\ee
which amounts, after simplifications, to
\begin{eqnarray}\label{eqh2c}
\fbox{  $\DD^b h^{(2)}_{ab}= \frac{1}{2} \DD^b k_{ac} k_{b}^{\:\: c}  + \DD_a \left( \sigma_c \sigma^c +8 \sigma^2 -\frac{1}{8} k_{cd} k^{cd}  +k_{cd} \sigma^{cd} \right) $ }
\end{eqnarray}
The radial equation of motion can be obtained after a straightforward, although tedious, computation and we find the quite intricate form
\beqs\label{eqh2d}
&\fbox{ $(\Box - 2) h^{(2)}_{ab} = 2 i_{ab} + \text{NL}_{ab}(\s,\s) + \text{NL}_{ab}(\s ,k)+\text{NL}_{ab}(k,k)$}
\eeqs
where the non-linear terms are given by
\beqs
\label{equation3bis}
\text{NL}_{ab}(\s ,\s )  &=& \DD_a \DD_b \left( 5 \sigma^2+\s_c \s^c  \right) + h_{ab}^{(0)} \left( -18 \s^2+4\s^c \s_c  \right) + 4 \s \s_{ab} \: , \nn \\
\text{NL}_{ab}(\s ,k)  &=& \DD_a \DD_b \left( k_{cd} \s^{cd}\right) -2k_{cd}\s^{cd}  h_{ab}^{(0)} + 4\s k_{ab}+4 \s^c (D_{(a}k_{b)c} - D_c k_{ab}) +4\s_{c(a} k^{c}_{\;\, b)} \: , \nn \\
\text{NL}_{ab}(k,k)  &=& k_{ac}k^c_{\; \, b} +k^{cd}( -\DD_{d}\DD_{(a}k_{b)c} +\DD_c \DD_d k_{ab}) \nonumber \\
&&  -\frac{1}{2}\DD_b k^{cd}\DD_a k_{cd}+\DD^d k_{c(a}\DD_{b)} k_{d}^{\; c}+\DD_c k_{ad}\DD^c k^d_{\; b}-\DD_c k_{ad}\DD^d k^{c}_{\; b}\, .\label{NLterms}
\eeqs
Using the relation $\s^c \s_{abc} = \s^c \s_{cab} + \s_a \s_b - h_{ab}^{(0)}\s_c \s^c$, one can rewrite the $\text{NL}_{ab}(\s ,\s )$ non-linear terms as
\be
\text{NL}_{ab}(\s ,\s ) = 6 \s_c \s^c h_{ab}^{(0)} + 8\s_a \s_b +14 \s \s_{ab} -18 \s^2 h_{ab}^{(0)}+2\s_{ac} \s^{c}_{\; b}+2\s_{abc}\s^c \, .
\ee
The equations of motion reproduce the expressions of \cite{BS} when $k_{ab}= i_{ab} = 0$.

\subsection{Equivalent systems for second order equations}
\label{equivsys}
Here, we would like to rewrite the second order equations in terms of two tensors  $V_{ab}$ and $W_{ab}$
\bea
V_{ab} &\equiv& - h_{ab}^{(2)}+\frac{1}{2} i_{ab}+ Q_{ab}^V \: ,\label{defnlVb} \\
W_{ab} &\equiv& \eps_a^{\;\, cd}D_c \left( h^{(2)}_{d b }-\frac{1}{2} i_{db} + Q_{db}^W \right) \: , \label{defnlWb}
\eea
where $Q_{ab}^{V,W}$ are appropriate quadratic terms in $(\s,\s)$, $(\s,k_{ab})$ or $(k_{ab},k_{ab})$ that we will construct herebelow. We require that the $V,W$ are SDT tensors that  obey the following duality properties
\bea
W_{ab} + \eps_a^{\;\, cd}D_c V_{d b } = K^W_{ab},\qquad V_{ab} - \eps_a^{\;\, cd}D_c W_{d b } = -2 i_{ab} + K^V_{ab}\, ,\label{nl03b}
\eea
where $K^{V,W}_{ab}$ are non-linear terms quadratic in $\sigma$ and $k_{ab}$ which are also SDT. Applying the curl operator on both equations \eqref{nl03b} we obtain that $V_{ab}$ and $W_{ab}$ obey
\bea
(\square - 2)V_{ab} = -2i_{ab} + K^V_{ab}+ \eps_a^{\;\, cd}D_c K^W_{d b } , \label{rhs01} \\
(\square - 2)W_{ab} = - 2 j_{ab} + K^W_{ab} - \eps_a^{\;\, cd}D_c K^V_{d b } ,  \label{rhs02}
\eea
where $j_{ab} \equiv -curl(i)_{ab}$. Our construction of the non-linear tensors $Q^{V,W}$, $K^{V,W}$ goes as follows. In order for $W_{ab}$ to be traceless, we require $Q^{W}_{ab}$ to be symmetric. Using the Hamiltonian and  momentum equation of motion, one can rewrite the symmetry condition of $W_{ab}$ as the following equation on $Q^{W}_{ab}$
\bea
\DD^b Q^W_{ab} - \DD_a Q^{W\, b}_b =  -\frac{1}{2}k^{bc} \DD_b k_{ac} + \DD_a \left( 4 \sigma^2 + \frac{3}{8} k_{cd} k^{cd} \right)  . \label{eqQ}
\eea
The divergence-free conditions of $W_{ab}$ can then be rewritten as
\bea
\eps_a^{\; \, cd} \DD_c \left( \DD^b Q^W_{d b} +\frac{1}{2} \DD_e k_{df} k^{ef} \right)= 0 \: ,
\eea
which is a consequence of the previous equation. The equation \eqref{eqQ} can be solved up to the ambiguity of adding to $Q^W_{ab}$ tensors obeying $\DD^b M_{ab} = \DD_a M$. We will fix the ambiguity in defining $Q_{ab}^W$ since we would like to find one equivalent formulation of the equations of motion, not all possible formulations. By choosing a $Q_{ab}^W$ with the smallest possible number of derivatives, we obtain
\bea
Q^W_{ab} = \left( - 2\sigma^2 +\frac{1}{16} k_{cd} k ^{cd} \right) h_{ab}^{(0)} - \frac{1}{2}k_{ac}k^{c}_{\; b}\, .\label{defQW}
\eea
Using again the Hamiltonian and momentum equation of motion, one can rewrite the traceless and divergence-free conditions of $V_{ab}$ as the following equations on $Q^{V}_{ab}$
\bea
Q_{a}^{V\; a} &=& 12 \sigma^2 + \sigma_c \sigma^c+\frac{1}{4} k_{cd} k^{cd}  + k_{cd} \sigma^{cd} \: , \\
\DD^b Q_{ab}^V &=& \frac{1}{2} \DD^b k_{ac} k_{b}^{\:\: c}  + \DD_a \left( \sigma_c \sigma^c +8 \sigma^2 -\frac{1}{8} k_{cd} k^{cd}  +k_{cd}\sigma^{cd} \right) \,.
\eea
This system has a unique solution up to the ambiguity of adding an SDT tensor to  $Q^{V}_{ab}$. We will do a definite choice for the ambiguity in defining $Q_{ab}^V$ as well. In fact, the SDT tensor $K^V_{ab}$ can be computed using \eqref{nl03b} and the equations of motion \eqref{equation3bis} as
\bea
K^V_{ab} &=& -NL_{ab}(\s,\s)-NL_{ab}(\s,k)-NL_{ab}(k,k) + Q^V_{ab} - (\Box - 3)Q^W_{ab} \nonumber \\
&& +\DD_a \DD^c (h^{(2)}_{bc} +Q^W_{bc})-h_{ab}^{(0)}(h^{(2)}+Q^{W \, a}_{a})
\eea
We will choose to fix the ambiguity  of adding SDT tensors to  $Q_{ab}^V$ by requiring
\bea
K^V_{ab} = 0 \: .\label{KVzero}
\eea
After a tedious computation, we obtain simply
\bea
Q_{ab}^V &=& (6\s^2 + \s^c \s_c+ \frac{1}{8} k_{cd} k^{cd}+\frac{1}{8} \DD_c k_{de} \DD^c k^{de} )  h_{ab}^{(0)} +2\s \s_{ab} -2\s_a \s_b \nonumber \\
&& + 4\:  \sigma k_{ab} - 4\:  \sigma^c \DD_c k_{ab} + 4 \:  \sigma^c \DD_{(a} k_{b)c} +  \:  4 \sigma_{c(a} k_{b)}^{\:\:\:c} - \:  \sigma_{cd} k^{cd} h^{(0)}_{ab} \nonumber\\
&& -\frac{1}{2} k_{ac} k_{b}^{\:\:c}  - \frac{3}{8} \DD_a k_{cd} \DD_b k^{cd} +\frac{1}{8} k^{cd} \DD_ {(a} \DD_{b)}  k_{cd} \, +Y^{(2)}_{ab} \: , \label{defQV}
\eea
where $Y^{(2)}_{ab}$ is an SDT tensor given in \eqref{Y2aa}.
%\bea
%Y^{(2)}_{ab}&\equiv&   \DD_{(a} k_{cd} \DD^c k_{b)}^{\:\:d}  - \DD_c k_{ad} \DD^d k_{b}^{\:\:c} - k^{cd} \DD_c \DD_{(a} k_{b)d}  + k^{cd} \DD_c \DD_d k_{ab}   \nonumber \\
 %&=& -4 \:  B^{(1)}_{c(a}   B^{(1) \: c}_{b)}  -2\: \epsilon_{cd(a}  \DD^c k_{b)}^{\;\:e}  B^{(1) \: d}_{e}  -2\: \epsilon_{cd(a}   \DD_{b)} B^{(1) \: c}_{e}   k^{de} \label{defNab}
%\eea
%is a SDT tensor which appeared in the classification in Appendix \ref{appSD}, see \eqref{defY2}.
Remark that to perform this computation, one can separate the analysis of non-linear terms for each set of quadratic terms $(k,k)$, $(\s,k)$ or $(k,k)$ independently since those terms never mix in the equations.

Using then the definition of $K^W_{ab}$ in \eqref{nl03b} we find
\bea
K^W_{ab} &=& curl(M)_{ab} =  \eps_a^{\;\, cd}  \DD_c M_{db} \: ,
\eea
where
\bea
M_{ab} \equiv Q^W_{ab}+Q^V_{ab}
\eea
is a tensor obeying $\DD^b M_{ab} = \DD_a M$. Using the classification of such tensors in Appendix \ref{sec:SD}, we have explicitly
\bea
M_{ab} = M^{[2,\s,\s,I]}_{ab}- M^{[2,\s,\s,II]}_{ab} + 4 M^{[2,\s,k]}_{ab} + Y^{(2)}_{ab}+ M^{[2,k,k,I]}_{ab}+ M^{[2,k,k,III]}_{ab} \: ,
\eea
where
\beqs
M^{[2,\s,\s, I]}_{ab}&=&  (5 \s^2 +\s_c \s^c ) h_{ab}^{(0)} + 4\s \s_{ab}, \nonumber \\
 M^{[2,\s,\s,II]}_{ab} &=&  (\DD_a \DD_b + h^{(0)}_{ab}) \: \s^2 ,  \nonumber \\
M_{ab}^{[2,\s,k]} &=&   \sigma k_{ab} - \:  \sigma^c \DD_c k_{ab} +  \:  \sigma^c \DD_{(a} k_{b)c} +  \:  \sigma_{c(a} k_{b)}^{\:\:\:c} -\frac{1}{4} \:  \sigma_{cd} k^{cd} h^{(0)}_{ab}   ,\nonumber \\
M^{[2,k,k,I]}_{ab}&=& \frac{1}{8}k_{cd} k^{cd} h_{ab}^{(0)} - k_{ac}k^c_{\; b} + \frac{1}{8}\DD_c k_{de}\DD^c k^{de} h_{ab}^{(0)}-\frac{1}{2}\DD_a k_{cd}\DD_b k^{cd} , \nonumber \\
M^{[2,k,k,III]}_{ab}&=& \frac{1}{16} \Big( \DD_a \DD_b + h^{(0)}_{ab}\Big)  k^{cd} k_{cd} \, .
\eeqs
In summary, the equations of motion can be written in the form:
\bea
W^a_a &=& \DD^b W_{ab}=0 \, ,\nonumber \\
(\square - 2)W_{ab} &=& curl(2 i + M )_{ab} \, ,\label{eqWfb}\\
i^a_a &=& \DD^b i_{ab} = 0 \, ,\nonumber\\
(\square -2)i_{ab} &=& 0\, .\label{eqib}
\eea
Using the curl operator, the first set of equations lead to
\bea
V^a_a &=& \DD^b V_{ab}=0 ,\nonumber \\
(\square - 2)V_{ab} &=& curl(curl(2 i + M ))_{ab} \, .\label{eqVfb}
\eea
Note that in the main text we have preferred to use the symmetrized curl of $\kappa_{ab}\equiv M_{ab}-M h^{(0)}_{ab}$. This is indeed equivalent to the curl of the tensor potential $M_{ab}$. One can also get rid of any SD tensor whose symmetrized curl is zero as a consequence of the properties of integration on the hyperboloid.

\section{Lemma on singular tensors}
\label{app:Nut}

This lemma was derived in collaboration with A. Virmani.

\begin{lemma}\label{newlemma} On the three dimensional hyperboloid, any scalar $\Phi$ satisfying $\square \Phi + 3\Phi = 0$ defines a symmetric, traceless, curl-free and divergence-free tensor $T_{ab} = D_a D_b \Phi + h^{(0)}_{ab}\Phi$ which can be written as
\begin{equation}
T_{ab} = \epsilon_{a}^{\;\, cd}D_c P_{d b},\label{eqkkk}
\end{equation}
where $P_{ab}$ is a symmetric, traceless tensor of the form
\bea
P_{ab} = \sum_{\mu=0}^3 N_{(\mu)} k^{(\mu)}_{ab} + P_{ab}^{reg},
\eea
where $P_{ab}^{reg}$ is regular and $k^{(\mu)}_{ab}$ are four singular tensors listed here below.
\end{lemma}
The regular tensors $P_{ab}^{reg}$ can be deduced from a Lemma stated in \cite{BS} and referred to as Lemma 2 in \cite{Compere:2011db}.
The four singular tensors $k_{ab}^{(\mu)}$ can be derived by integrating equation \eqref{eqkkk} for $\Phi = \hat \zeta_{(\mu)}$ where $\hat \zeta_{(\mu)}$ are the four solutions of $(\square +3)\Phi = 0$ which are even under parity-time reversal and contain the harmonics $l=0$ or $l=1$ on the two-sphere. Explicitly,
\bea
\hat \zeta_{(0)} &=& \frac{\cosh 2\tau}{\cosh\tau}, \qquad \hat \zeta_{(1)} = \left( 2 \sinh\tau + \frac{\tanh \tau }{\cosh \tau} \right) \cos\theta ,\\
\hat \zeta_{(2)} &=& \left( 2 \sinh\tau + \frac{\tanh \tau }{\cosh \tau} \right) \sin\theta \cos\phi \qquad \hat \zeta_{(3)} = \left( 2 \sinh\tau + \frac{\tanh \tau }{\cosh \tau} \right) \sin\theta \sin\phi \,.
\eea
The four singular tensors can be written in the traceless gauge $h_{(0)}^{ab}k_{(\mu)ab} = 0$ as
\begin{eqnarray}
k_{(0)ab} &=& \left( \begin{array}{ccc}  0&0 & 2 \frac{{\hat k}-\cos\theta}{\cosh\tau}\\ 0 &0 & \sinh\tau \frac{\cos{2\theta}-4{\hat k} \cos\theta +3}{2\sin\theta} \\ 2 \frac{{\hat k}-\cos\theta}{\cosh\tau} &  \sinh\tau \frac{\cos{2\theta}-4{\hat k} \cos\theta +3}{2\sin\theta}&0 \end{array}\right),\nonumber \\
k_{(1)ab} &= &\left( \begin{array}{ccc}  0&0 & -3 \frac{\tanh\tau}{\cosh\tau}\sin^2\theta\\ 0 &0 & \frac{-8{\hat k}+9\cos\theta - \cos 3\theta}{4\sin\theta} \cosh\tau\\ -3 \frac{\tanh\tau}{\cosh\tau}\sin^2\theta &  \frac{-8{\hat k}+9\cos\theta - \cos 3\theta}{4\sin\theta} \cosh\tau  &0 \end{array}\right),\label{kiab} \\
k_{(2)ab} &= &\left( \begin{array}{ccc} 0& 3 \frac{\tanh\tau}{\cosh\tau}\sin\phi   & 3 \frac{\tanh\tau}{\cosh\tau}\cos\theta \sin\theta \cos\phi \\ 3 \frac{\tanh\tau}{\cosh\tau}\sin\phi & \frac{8{\hat k}-9\cos\theta+\cos 3\theta}{2\sin^3\theta } \cosh\tau \sin\phi & \frac{\cos^4\theta-4{\hat k} \cos\theta +3}{\sin^2\theta}\cosh\tau\cos\phi \\ 3 \frac{\tanh\tau}{\cosh\tau}\cos\theta \sin\theta \cos\phi &   \frac{\cos^4\theta-4{\hat k} \cos\theta +3}{\sin^2\theta}\cosh\tau\cos\phi & \frac{-8{\hat k}+9\cos\theta-\cos 3 \theta}{2\sin\theta}\cosh\tau \sin\phi \end{array}\right),\nonumber \\
k_{(3)ab} &= &\left( \begin{array}{ccc} 0& -3 \frac{\tanh\tau}{\cosh\tau}\cos\phi   & 3 \frac{\tanh\tau}{\cosh\tau}\cos\theta \sin\theta \sin\phi \\ -3 \frac{\tanh\tau}{\cosh\tau}\cos\phi & \frac{-8{\hat k}+9\cos\theta-\cos 3\theta}{2\sin^3\theta } \cosh\tau \cos\phi & \frac{\cos^4\theta-4{\hat k} \cos\theta +3}{\sin^2\theta}\cosh\tau\sin\phi \\ 3 \frac{\tanh\tau}{\cosh\tau}\cos\theta \sin\theta \sin\phi &   \frac{\cos^4\theta-4{\hat k} \cos\theta +3}{\sin^2\theta}\cosh\tau\sin\phi & \frac{8{\hat k}-9\cos\theta+\cos 3 \theta}{2\sin\theta}\cosh\tau \cos\phi \end{array}\right).\nonumber
\end{eqnarray}
These tensors are regular in the north patch upon choosing ${\hat k} = +1$ and in the south patch upon choosing ${\hat k} = -1$. They are tranverse and obey the equation
\bea
(\square -3)k_{(\mu) ab} = 0
\eea
outside of singularities. The singular transition function between the south and north patches can be written as

\begin{eqnarray}
\delta k_{(0)ab} \equiv k_{(0)ab}|_{South} - k_{(0)ab}|_{North}  &=& \left( \begin{array}{ccc}  0&0 & -\frac{4}{\cosh\tau} \\ 0&0 &4\cot\theta \sinh\tau \\ -\frac{4}{\cosh\tau} & 4\cot\theta \sinh\tau & 0 \end{array}\right),\nonumber \\
\delta k_{(1)ab} \equiv k_{(1)ab}|_{South} - k_{(1)ab}|_{North}  &=& \left( \begin{array}{ccc} 0 &0 & 0\\ 0&0 & 4\frac{\cosh\tau}{\sin\theta} \\ 0& 4\frac{\cosh\tau}{\sin\theta}& 0\end{array}\right),\nonumber \\
\delta k_{(2)ab} \equiv k_{(2)ab}|_{South} - k_{(2)ab}|_{North}  &=& \left( \begin{array}{ccc} 0 &0 &0 \\0 & -\frac{8}{\sin^3\theta}\cosh\tau \sin\phi & \frac{8 \cos\theta}{\sin^2\theta}\cosh\tau \cos\phi \\ 0&\frac{8 \cos\theta}{\sin^2\theta}\cosh\tau \cos\phi  &  \frac{8}{\sin\theta} \cosh\tau \sin\phi \end{array}\right),\nn \\
\delta k_{(3)ab} \equiv k_{(3)ab}|_{South} - k_{(3)ab}|_{North}  &=& \left( \begin{array}{ccc} 0 &0 &0 \\0 & \frac{8}{\sin^3\theta}\cosh\tau \cos\phi & \frac{8 \cos\theta}{\sin^2\theta}\cosh\tau \sin\phi \\ 0&\frac{8 \cos\theta}{\sin^2\theta}\cosh\tau \sin\phi  &  -\frac{8}{\sin\theta} \cosh\tau \cos\phi \end{array}\right).\nonumber
\end{eqnarray}
These transition functions obey
\begin{eqnarray}
D_{[a} \delta k_{(\mu) b]c} = 0, \qquad (\square - 3) \delta k_{(\mu)ab} = 0,\qquad  h^{(0)\: ab}\delta k_{(\mu)ab}=0,\qquad D^b \delta k_{(\mu)ab}=0, \label{propdeltak}
\end{eqnarray}
on the hyperboloid outside the singular region $\theta =0$ and $\theta =\pi$ and obey the normalized orthogonality relations
\bea
\int_0^{2\pi} d\phi \;\delta k_{(\mu) \phi a}D^a \zeta_{(\nu)} = - 8\pi \; \delta_{(\mu)(\nu)},\qquad \mu,\nu=0,\dots 3\, ,\label{orthok}
\eea
where $\zeta_{(\mu)}$ are the four solutions of $\DD_a \DD_b \zeta_{(\mu)} + h_{ab}^{(0)}\zeta_{(\mu)}=0$ given by
\bea
\zeta_{(0)} &=& - \sinh \tau ,\qquad \zeta_{(1)} = \cosh\tau \cos\theta, \\
\zeta_{(2)} &=& \cosh\tau \sin\theta \cos\phi , \qquad \zeta_{(3)} = \cosh\tau \sin\theta \sin\phi
\eea
which are odd under parity-time reversal and normalized such that $\zeta_{(\mu)}\p_\rho + \rho^{-1}\p^a \zeta_{(\mu)} \p_a +o(\rho^{-1})= \p_\mu$ where $\p_\mu = \p_t,\, \p_i$.

\end{document}